# A graph neural network surrogate model for mesh-based crashworthiness prediction of vehicle panel components

**Haoran Li [1], Yingxue Zhao [1], Haosu Zhou [1], Tobias Pfaff [2], Nan Li [1*]**

[1] Dyson School of Design Engineering, Imperial College London, London, UK

[2] NVIDIA, UK

* Corresponding author. E-mail address: n.li09@imperial.ac.uk (N. Li)

## Abstract

Crashworthiness is a key performance measure in the design of safety-critical vehicle panel components such as B-pillars. Finite element (FE) simulations are widely used to evaluate crash responses but remain computationally expensive for large-scale, nonlinear impact scenarios, particularly when integrated into iterative design and optimisation processes. Although machine learning-based surrogate models have been developed for rapid crashworthiness analysis, they exhibit limitations in detailed representation of complex 3-dimensional components. Graph Neural Networks (GNNs) have emerged as a promising solution for processing data with complex structures. However, existing GNN models often lack sufficient accuracy and computational efficiency to meet industrial demands. This paper proposes Recurrent Graph U-Net (ReGUNet), a graph-based surrogate model for crashworthiness analysis of vehicle panel components. By representing FE meshes in graph form, the model naturally accommodates complex irregular structural geometries. Its hierarchical architecture improves computational efficiency and accuracy, while the introduction of recurrence enhances stability of temporal predictions over multiple time steps. A side-impact case study of hot-stamped steel B-pillars with varying geometries is used to generate training dataset. The trained model demonstrates high accuracy in predicting the dynamic deformation behaviour and crashworthiness indicators of previously unseen component designs. ReGUNet achieves over a 52% reduction in the average deformation prediction error relative to baseline methods, together with markedly improved computational efficiency. ReGUNet provides rapid and reliable crashworthiness assessments, which in turn accelerates the design cycle of vehicle panel components.

**Keywords**: Artificial intelligence; Surrogate modelling; Deep learning; Graph neural network; Crashworthiness analysis

# 1 Introduction

Vehicle lightweighting has been a central focus in the automotive industry due to the growing concern about global climate change [1]. Significant efforts are devoted to developing innovative design methodologies for vehicle components that achieve weight reduction while ensuring high structural performance. For vehicle safety-critical panel components like the A-pillar, B-pillar, and front roof cross member, crashworthiness performance remains a primary design requirement. This measures the component's ability to protect passengers in potential vehicle accidents, including deformation resistance and energy absorption during collision. Typical crashworthiness analysis of a vehicle component involves studying the dynamic behaviour of the component during crash tests. For example, side impact crash tests assess lateral collision performance, in which the B-pillar is a key thin-walled structural element [2]. Common crashworthiness indicators include maximum B-pillar intrusion and energy absorption. Although physical vehicle crash tests provide reliable insight into component crashworthiness performance, they are too expensive to be used iteratively in design cycles. Finite element (FE) simulations serve as a computational alternative. However, they require significant computational resources for complex and highly nonlinear scenarios, particularly under conditions involving large deformation and high strain rates. Furthermore, iterative design optimisation processes that rely heavily on FE simulations can be excessively time-consuming.

To overcome these limitations, machine learning (ML)-based surrogate modelling has been widely explored to approximate crash simulations with significantly reduced computational cost. The surrogate models are usually constructed using artificial neural networks (ANNs), which aim to approximate complex simulations with reduced computational resource consumption. The application of ANN-based surrogate models has shown potential within the area of crashworthiness. Most existing surrogate models [3-8] were based on multilayer perceptron (MLP) and were designed for the prediction of scalar quantities such as the crashworthiness indicators. A few studies [9, 10] used recurrent neural networks (RNNs) for time series data predictions, which are essential for crashworthiness analysis. In time series data, numerous measurements, such as acceleration profiles and force-time histories, are recorded over specific time intervals. Both MLP and RNN-based models are restricted to predictions based on scalar inputs and outputs, they exhibit limitations in fully describing the detailed behaviours of complex simulations, such as collision modes. To overcome this limitation, field-based models developed using convolutional neural networks (CNNs) have been proposed to predict high-fidelity physical fields, such as displacement fields during impact [11, 12]. Field-based models enable more comprehensive and detailed analyses of component behaviours during impact. However, typical crashworthiness studies involve studying nonlinear dynamic behaviours of thin-walled structures with irregular geometries and unstructured FE meshes. CNN-based surrogate models are inherently limited to Euclidean data structures, such as images. They have limited capabilities of processing complex geometries characterised by non-Euclidean data structures. This limitation highlights the importance of the development of more advanced surrogate models that can process more complex data structures while maintaining a high level of accuracy.

Graph neural networks (GNNs) provide a natural framework for addressing this limitation due to direct graph representation form of 3-dimensional structures. Wen et al. [13] utilised GNNs to predict dynamic behaviours of vehicle crashworthiness components. In this work, vehicle components were represented in terms of graphs, where each node represents a segment and each edge describes the connectivity between segments. Despite the strong predictive performance demonstrated on regularly structured components, the approach's ability to handle complex irregular-shaped components remain unexplored. A more flexible approach is to represent FE meshes as graphs, with graph nodes corresponding to mesh nodes and graph edges capturing connectivity. Recent studies have utilised GNNs in general mechanics-related domains such as continuum solid mechanics [14-18], structural mechanics [19, 20], computational fluid dynamics [21-24], and metamaterial modelling [25, 26]. Some pioneering works have applied GNNs to Lagrangian mesh-based simulations [27-29], but these studies have generally been restricted to simplified loading conditions with small graphs. Although these studies established a strong foundation for GNN applications in mechanics-related domains, their ability to handle industrially relevant crash scenarios, involving large graphs, complex impact conditions, and nonlinear dynamic behaviour, remains largely unexplored. To the best of the authors' knowledge, mesh-based GNN surrogate model has not yet been applied to crashworthiness studies.

In this paper, we propose a new GNN-based surrogate model, namely the Recurrent Graph U-Net (ReGUNet), to address these challenges in crashworthiness prediction. ReGUNet is designed for spatiotemporal full-field prediction of dynamic deformation of thin-walled automotive components under impact loading. By integrating hierarchical graph coarsening with a recurrent temporal mechanism, ReGUNet achieves efficient long-range spatial information propagation and stable temporal predictions, tailored to the nonlinear, high strain-rate and large-deformation characteristics of crash problems. To demonstrate its effectiveness, ReGUNet is evaluated on a side-impact case study of hot-stamped steel B-pillars with varying geometries. The results show that ReGUNet provides accurate and efficient predictions of crash responses, offering a practical tool to accelerate the design and optimisation of thin-walled automotive structures.

This paper first reviews relevant existing GNN applications in crashworthiness studies and related fields in Section 2. This is followed by introducing the crashworthiness case study, a B-pillar side impact test, in Section 3. The details of ReGUNet are described in Section 4, including explaining the graph representation methods and the neural network architecture. The evaluation of the performance of ReGUNet is discussed in Section 5. This is followed by concluding the current work and stating the future directions in Section 6. The primary objectives of this work are:

- To propose a GNN-based surrogate model, Recurrent Graph U-Net (ReGUNet), for rapid prediction of the dynamic crash response of vehicle panel components.
- To construct a B-pillar side-impact simulation dataset via mesh morphing, enabling supervised training and evaluation on node-level deformation fields and graph-level quantities such as energy absorption and contact forces.
- To perform a comparative assessment of ReGUNet against representative baseline surrogate models and ablated variants, including sensitivity to key architectural design choices.
- To quantify the predictive accuracy and computational efficiency of the proposed surrogate with respect to high-fidelity FE simulations.

# 2 Related work

To the best of the author's knowledge, the application of GNNs in crashworthiness analysis is still in its early stages. To date, only a limited number of studies have explored GNN-based approaches for modelling the complex dynamics of vehicle components under impact conditions. One notable example is the work by Wen et al. [13], who proposed a model combining GNN with Temporal Convolutional Neural Networks (TCN) for predicting the nonlinear response of vehicle components represented as graphs, where nodes denote component segments and edges denote adjacency. While the nodes and edges determine the geometry of the component, the node features contain information on measurements of crashworthiness analysis. The model employs an encoder–decoder spatiotemporal GNN architecture. Specifically, the spatiotemporal GNN layers consist of gated temporal convolutional layers [30] for temporal modelling and graph convolutional network [31] layers capturing spatial information. Nodes and edge features, together with the global parameters such as initial impact velocity are input into the model to predict the updated node features for the following time step. One limitation of the aforementioned model is that it has been validated only on regularly structured components, leaving its performance on irregular graph data unverified.

A number of existing studies have employed GNNs on mesh-based data prediction in related engineering domains, such as structural mechanics, fluid dynamics, and material modelling [32]. In most studies, mesh data is defined in terms of graph, where each graph node represents a mesh node, and each graph edge represents a mesh edge. For instance, Deshpande et al. [14] proposed the Multi-channel Aggregation Network (MAgNET) for the prediction of non-linear mesh-based simulations. The MAgNET consists of Multi-channel Aggregation (MAg) layers which assign a non-shareable weight to each edge of the graph for local feature aggregation. Unlike conventional CNNs using sharable weights (sliding kernels), the non-shareable weight matrix allows more accurate non-linear feature prediction across different channels. A number of other studies utilised MLPs for feature updating and aggregation, which can be more generalisable and scalable. Sanchez-Gonzalez et al. [33] developed the Graph Network-based Simulators (GNS) for predicting the motions of physical systems with particles. GNS follows an encoder-processor-decoder architecture. The framework is based on a sequence of Graph Network (GN) blocks to achieve long range message passing across graph data. Pfaff et al. [27] adopted and improved this framework by proposing the MeshGraphNet (MGN). MGN was specifically designed for mesh-based data prediction utilising message passing neural network [34]. The model introduces extra world-edges in addition to the mesh-edges to achieve message passing across different graphs to better simulate collisions. To perform node-level prediction task, the model uses MLPs as edge and node update functions.

To achieve spatially distant information propagation, the conventional approach is to stack a sequence of GN blocks that allows multi-step message passing. This usually results in a significant increase in GPU memory consumption when processing large graphs. To address this limitation, Fortunato et al. [29] proposed a hierarchical architecture, the MultiScale MeshGraphNet (MS-MGN), which performs message passing across graphs with different resolutions. Higher efficiency is achieved due to the reduced number of message passing steps required from one global location to another in the low-resolution graph. Information is propagated between graphs through downsampling and upsampling layers. Graphs with different resolutions are constructed based on manually produced meshes. Cao et al. [28] proposed an alternative downsampling approach with their Bi-Stride Multi-Scale GNN (BSMS-GNN). In this study, a novel bi-stride downsampling technique was introduced based on the breadth-first search (BFS). The graph size can be reduced by striding and pooling all nodes at every other BFS frontier. There are other graph downsampling approaches such as the spatial proximity approach [35-37], Guillard's coarsening algorithm [38], and the clustering-based pooling method [14]. All these different approaches serve the same purpose of reducing the complexity of graph data, hence improving model's efficiency. One limitation of these automated graph downsampling approaches is the potential loss of uniformity in the downsampled graphs, which may adversely impact the accuracy of message-passing operations.

Common GNN tasks in mechanics often involve temporal data prediction, where the objective is to predict feature evolution over time. For example, Pfaff et al. [27] used MGN to predict mesh deformation of metal plates under quasistatic loading. The deformation is divided into certain time intervals, and each training iteration predicts the incremental deformation during each time interval. The prediction of the next immediate step is based on the previous steps, resulting in an increase in rollout error when the number of time steps increases. Chen et al. [17] proposed the physics-informed edge recurrent simulator (Piers) to improve this limitation. Piers was developed based on MGN, with the incorporation of RNN and physics-informed factors. In the GNN cell, the edge update function is replaced with a gated recurrent unit (GRU) layer to better capture long-horizon dependencies. This results in a reduction in rollout error when the number of time steps increases and leads to a more stable prediction.

Piers is an example of a broader class of physics-informed surrogate models, where knowledges such as constitutive laws or energy consistency are embedded into the learning objective or the architecture [17]. In solid mechanics, physics-informed GNNs have been proposed for soft-tissue and hyperelastic materials, where mesh-based GNNs are trained via minimisation of total potential energy or deep-energy functionals, leading to more physically consistent deformation fields than purely data-driven surrogates [39, 40]. More generally, physics-consistent neural networks enforce invariants such as energy or mass conservation via output projections or soft penalty terms, which has been shown to enhance robustness and interpretability of surrogates for dynamical systems[41].

Despite the achievements of the aforementioned GNN applications, most models have only been evaluated on relatively simple case studies with small-scale graphs. In the solid mechanics field, MGN and its extended models are capable of predicting plate deformation under quasistatic loading conditions. High-speed dynamic systems are often more complex as they become highly nonlinear. Piers was designed for the prediction of dynamic responses of continuous deformable bodies (CDBs) with nonlinearity [17]. However, the graph size is limited as the GPU memory consumption would be substantial with large graphs. For crashworthiness analysis of vehicle panel components, high-speed impact and fine mesh configuration are inevitable. Surrogate model training typically involves hundreds of samples with varying geometric designs, therefore, model efficiency becomes a crucial factor to consider when designing the architecture. In this paper, we propose the Recurrent Graph U-Net (ReGUNet), which is specifically tailored to overcome the aforementioned limitations.

# 3 Simulation setup and dataset generation

This section describes the FE simulation setup used to generate the dataset for model training and evaluation. A hot-stamped steel B-pillar subjected to a side-impact scenario is analysed. The FE model is developed to represent a simplified crash test focused solely on the B-pillar at the component level [44]. As illustrated in Figure 1 a), the B-pillar is fixed at its upper and lower extremities, with a hemispherical impactor positioned to strike at the lower region, simulating a vehicle side crash scenario. Figure 1 b) shows the FE setup of the impact test, fixed boundary conditions in all six degrees of freedom are applied to the top and bottom ends of the B-pillar, and the impactor collides with the component with an initial velocity in the negative z direction. The resultant displacement fields are extracted, which depict the deformation during impact. Figure 1 c) shows an example of the z-displacement field.

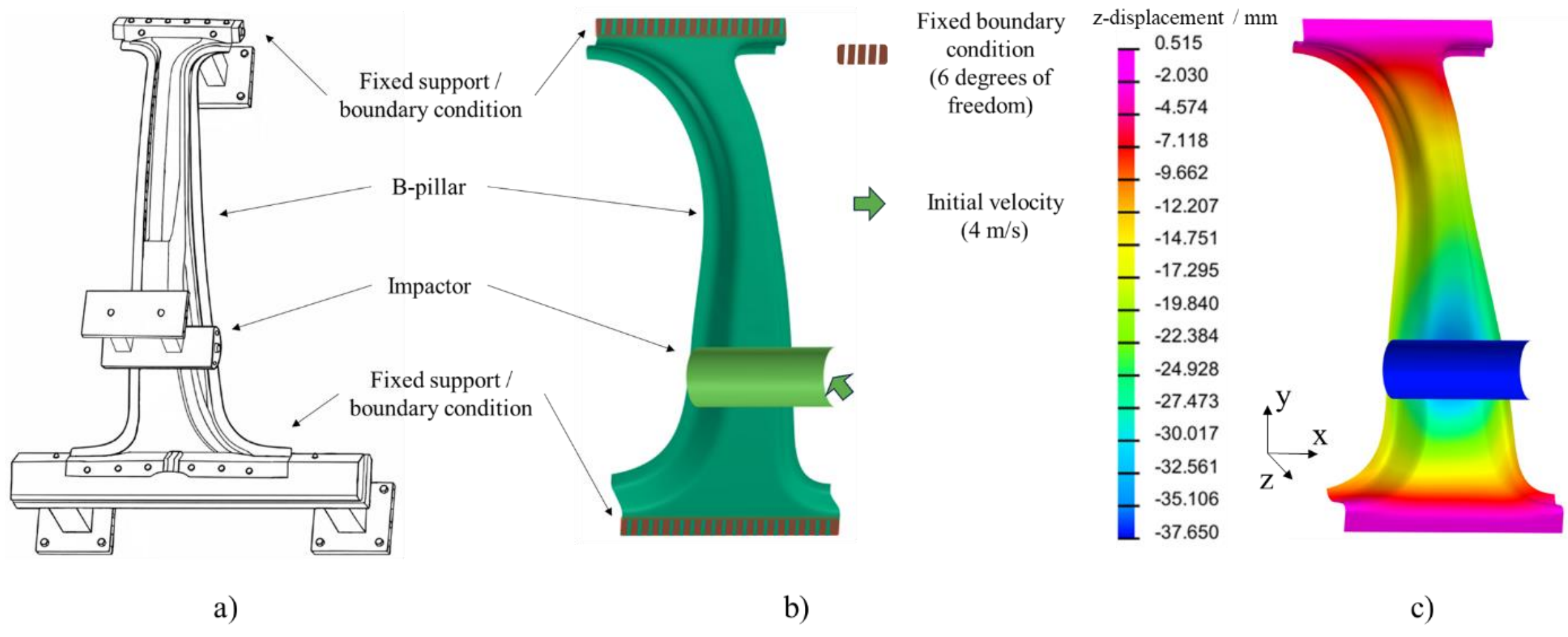

Figure 1: Illustration of a) experimental setup of the B-pillar impact test as described in [44], b) FE setup of the B-pillar impact test, and c) result contour of the FE simulation showing the z-displacement in mm.

In order to study the influence of designed shape on the crashworthiness performance, the material properties, boundary and loading conditions remain constant throughout all samples. Component shape variation is achieved via mesh morphing. For the loading condition, the initial position of the impactor is adjusted on top of the geometry without any offset according to the mesh morphing to ensure constant loading condition. The FE model of the B-pillar is a scaled-down representation, approximately one-third the size of a full-scale B-pillar component used in real-world applications. Correspondingly, the impactor is proportionally scaled down by the same factor, maintaining the same size ratio. The length of the B-pillar is 350 mm, the impactor has a diameter of 40 mm and a mass of 100 kg, and the initial impact velocity is 4 m/s.

The simulations are conducted using Virtual-Performance Solution (VPS-2022) with an explicit time integration scheme. The B-pillar is modelled using triangular shell elements with uniform thickness of 1.6 mm and an average element size of 3 mm, resulting in approximately 4000 nodes and 7,000 to 8,000 shell elements per model. Hourglass control is employed in the FE simulations to suppress zero-energy deformation modes inherent to reduced-integration shell elements, ensuring numerical stability. Contact between the impactor and B-pillar is defined using a surface-to-surface contact algorithm (type 33) with a friction coefficient of 0.1. The simulation duration is 30 ms, with stable time steps determined automatically by the solver. When constructing graph sequences from the dynamic simulations, the entire crash duration is divided into 20 time steps, with each step separated by an equally spaced time interval of 1.5 ms. However, only the first 12 steps are used for model training and evaluation, as the B-pillar undergoes no further deformation beyond this point. This allows for consistent temporal analysis throughout the simulation, ensuring that the progression of the deformation can be accurately modelled and captured at uniform time intervals. The details for graph sequence construction are described in Section 4.1.

The material of the B-pillar is defined to be boron steel with the 100% martensite phase under room temperature, which is a strain-rate dependent material. Its property is defined with the constitutive model developed by Li [45], then input into VPS as look-up table. Figure 2 shows the relationship between true stress and true strain of the material at four representative strain rates. Damage evolution is not included in the present material model. This is a deliberate modelling choice because this surrogate modelling methodology proposed in this work is intended to aid early-stage design workflows by enabling fast, mesh-aware full-field prediction. Such simplified material descriptions are often used in early-stage design and method development. In practice, explicit damage or fracture formulations are typically introduced in later-stage assessments or failure-dominated studies, because they require extensive calibration and add substantial computational overhead. Importantly, the proposed framework is extensible and can be augmented in future work to learn from damage-coupled simulations or incorporate damage-related criteria.

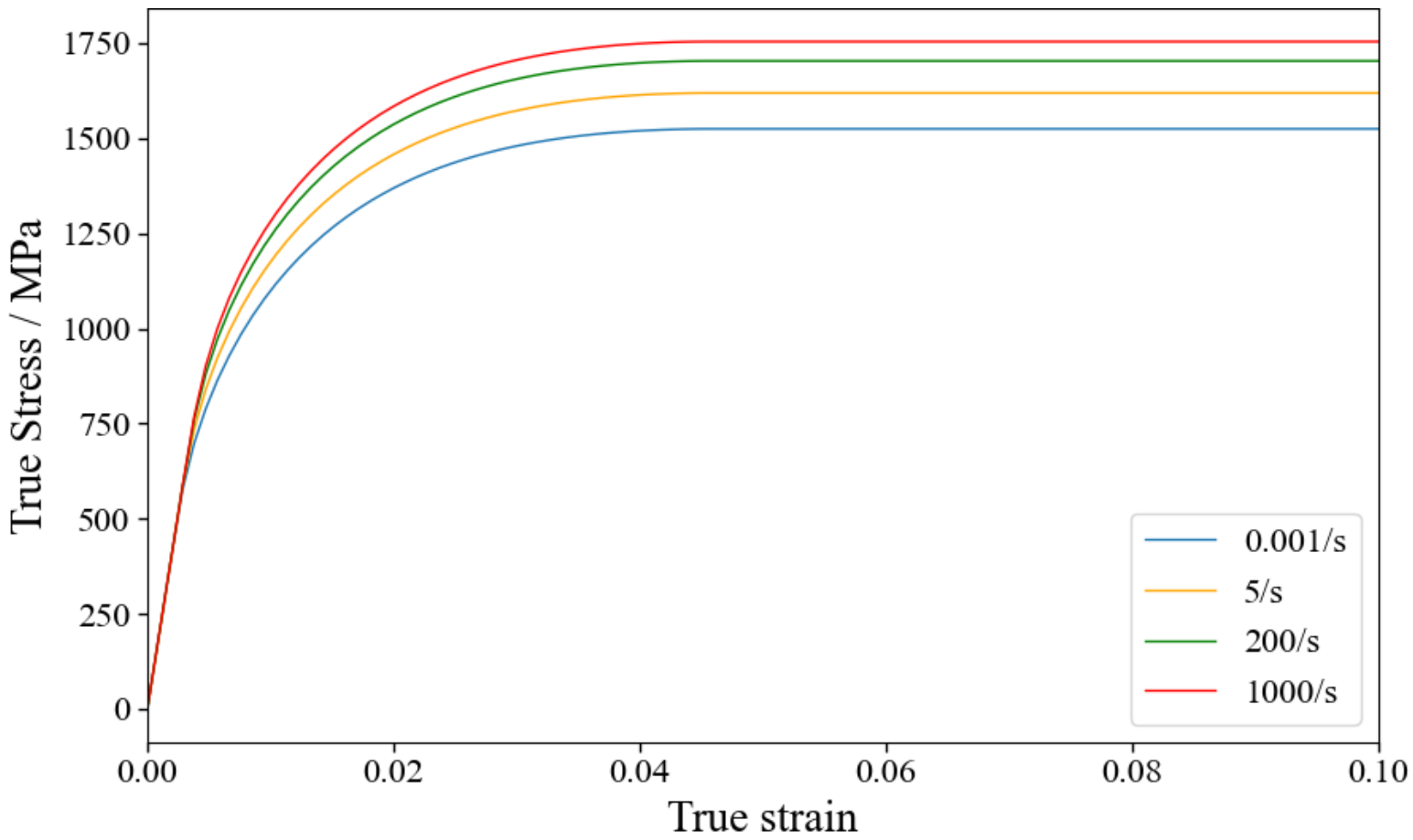

Figure 2: Material properties of the boron steel with different strain rates under room temperature in terms of true stress vs. true strain.

The design modification is made through mesh morphing, which alters the shape of the component by directly manipulating its geometry. As shown in Figure 3, the B-pillar is morphed in the z direction at one of the three control points defined on the component. The morphing distance is controlled to remain within ±8% of the total thickness in the z-direction of the baseline B pillar. Optionally, the top region of the component is morphed in the x direction to the right within 10% of the total component width. Latin hypercube sampling (LHS) is used to randomly sample both the control point variable and the morphing distance, ensuring a comprehensive exploration of the design space while maintaining a uniform distribution of the sampled variables. This morphing strategy is used to generate two case studies. Case study 1 is simpler as the component is only morphed in the z direction at one control point at a time. In this case study, only node-level outputs are predicted, which are the x, y, and z displacement fields. This allows the model to learn how the curvature of the B-pillar affects the deformation during crash tests. Case study 2 is more complex, involving simultaneously morphing in both x and z directions. This makes the data covers a much wider design space of shape. In addition, the model not only predict nodal displacement fields, but also outputs graph-level scalar predictions for other engineering key performance indicator (KPIs). This includes the component's internal energy (estimation of energy absorption during crash test) and the contact force over time. For both case studies, three sets of data are generated with separate LHS operations for training, validation, and test, each covering the entire design space. This ensures that every dataset has diverse and representative samples from the design space. The training set contains 100 samples, while the validation and test sets each consists of 50 samples. Each sample consists of a graph sequence of 12 graphs including the initial state, representing the intermediate frames at each time step throughout the crash simulation. The diversity of the sampled dataset is evaluated in Appendix A.

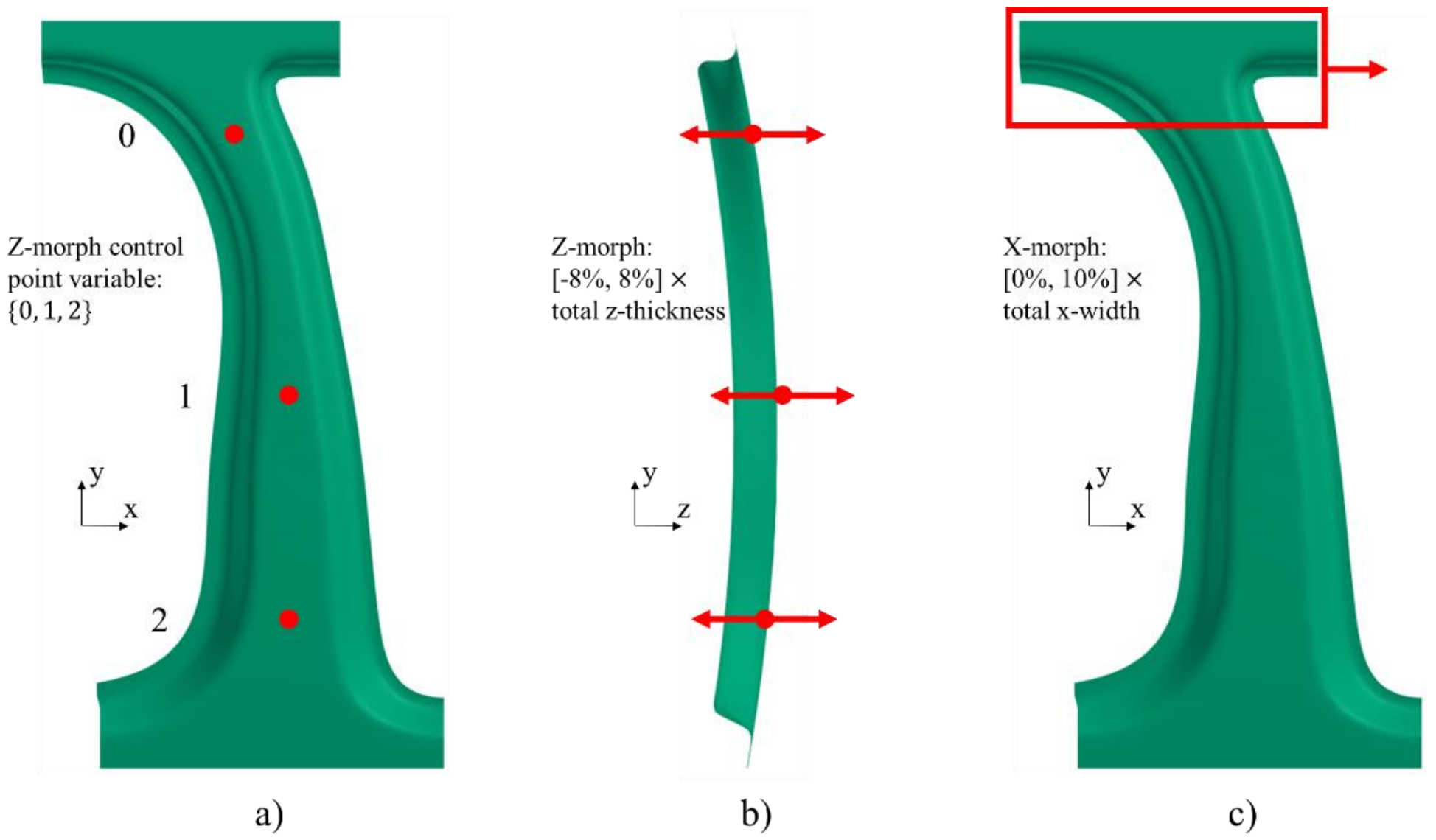

Figure 3: Morphing variables of the B-pillar component. a) the z morphing control points, b) the z morphing ranges, and c) the x morphing region and range.

# 4 The development of Recurrent Graph U-Net

This section presents a detailed description of the Recurrent Graph U-Net (ReGUNet). We first introduce the graph representation form of FE meshes, followed by an explanation of the model architecture. The model aims to predict the full impact dynamics of the B-pillar given varying geometry designs.

## 4.1 Graph representation of FE meshes

A typical graph $G$ can be expressed by $G = (\mathbf{u}, V, E)$, where $\mathbf{u}$ is the global graph feature, $V = \{v_1, v_2, \dots, v_n, \dots, v_N\}$ is a set of nodes and $E = \{e_1, e_2, \dots, e_m, \dots, e_M\}$ is a set of edges connecting nodes. $N$ and $M$ are the numbers of nodes and edges, respectively. The connectivity of a graph describes how the nodes are connected by edges, which can be represented by the edge index. An edge index matrix $E_{index}$ is a $2 \times M$ matrix containing the node indices connected by each edge, and can be represented as

$$E_{index} = \begin{bmatrix} v_{s_1} & v_{s_2} \dots v_{s_M} \\ v_{r_1} & v_{r_2} \dots v_{r_M} \end{bmatrix}, \tag{1}$$

where $v_s$ and $v_r$ are the sender and receiver nodes of each edge, respectively. On this graph, data can be stored by locally embedding it in the node feature vector $\mathbf{x}_{v_n} \in \mathbb{R}^{p_v}$ and the edge feature vector $\mathbf{x}_{e_m} \in \mathbb{R}^{p_e}$, where $p_v$ and $p_e$ are the dimensions of the node and edge feature vectors, respectively. Global graph feature $\mathbf{u} \in \mathbb{R}^{p_u}$ includes any global scalar features, where $p_u$ is the dimension of the graph feature vector. The assembled matrices $\mathbf{X}_V \in \mathbb{R}^{N \times p_v}$ and $\mathbf{X}_E \in \mathbb{R}^{M \times p_e}$ denote the node and edge feature matrices, respectively. In a deep GNN, multiple layers are typically employed. The features are aggregated and propagated through each layer, where $\mathbf{X}^{(l)}$ and $\mathbf{x}^{(l)}$ denote the hidden node or edge feature matrix and vector at the $l$-th layer.

The graph data representation of the FE mesh of a B-pillar is illustrated in Figure 4. Similar to the existing methods [17, 27, 29], each graph node represents an FE mesh node, and edges connect any pairs of nodes that belong to the same mesh element. In order to predict the impact dynamics of the component, the crash simulation is divided into equally spaced time intervals. The temporal evolution of crash dynamics is thus described by a sequence of graphs $G_{seq} = [G^{t_1}, G^{t_2}, \dots, G^{t_T}]$, where $T$ is the total number of time steps. As no mesh refinement is applied to the simulations, the graph connectivity stays constant throughout all time steps. The node and edge features, which represent the physics-field information of interest and change dynamically with time, describe the current state of the graph.

We encode graph features as follows: The node features are defined based on the incremental displacements of each mesh node. Specifically, for a given node $v_n$ at time step $t_i$ , the node feature vector is defined as: $\mathbf{x}_{v_n}^{t_i} = (s_{n_x}^{t_i}, s_{n_y}^{t_i}, s_{n_z}^{t_i})$ , where $s_{n_x}^{t_i}$ , $s_{n_y}^{t_i}$ , and $s_{n_z}^{t_i}$ represent the incremental displacements in the x, y, and z directions, respectively, from the previous time step $t_{i-1}$ to the current $t_i$. These components are concatenated into a 3-dimensional feature vector for each node, serving as input to the GNN model. In order to achieve translation-invariance, absolute nodal coordinates are not included in the node features. Rather, we encode the geometric features using relative distances between nodes and include them into edge features. Specifically, for an edge $e_m$, the edge feature vector at time $t_i$ consists of 8 components, $\mathbf{x}_{e_m}^{t_i} = (d_{m_x}^{t_i}, d_{m_y}^{t_i}, d_{m_z}^{t_i}, |d_m^{t_i}|, d_{m_x}^{t_1}, d_{m_y}^{t_1}, d_{m_z}^{t_1}, |d_m^{t_1}|)$, where $d$ denotes the relative distances in the x, y, and z directions between the connected nodes and $|d|$ denotes the corresponding Euclidean distance. The edge feature vector includes information at both time $t_i$ and the initial time step $t_1$. This information describes both the component's current shape during impact and its original shape. Taking the input node and edge features, the model predicts an updated set of node features $\mathbf{x}_{v_n}^{t_{i+1}}$, namely the incremental displacements for each node at time interval $t_{i+1}$. The graph feature $\mathbf{u}$ is optional, which can include component-level scalar information such as total internal energy and contact force.

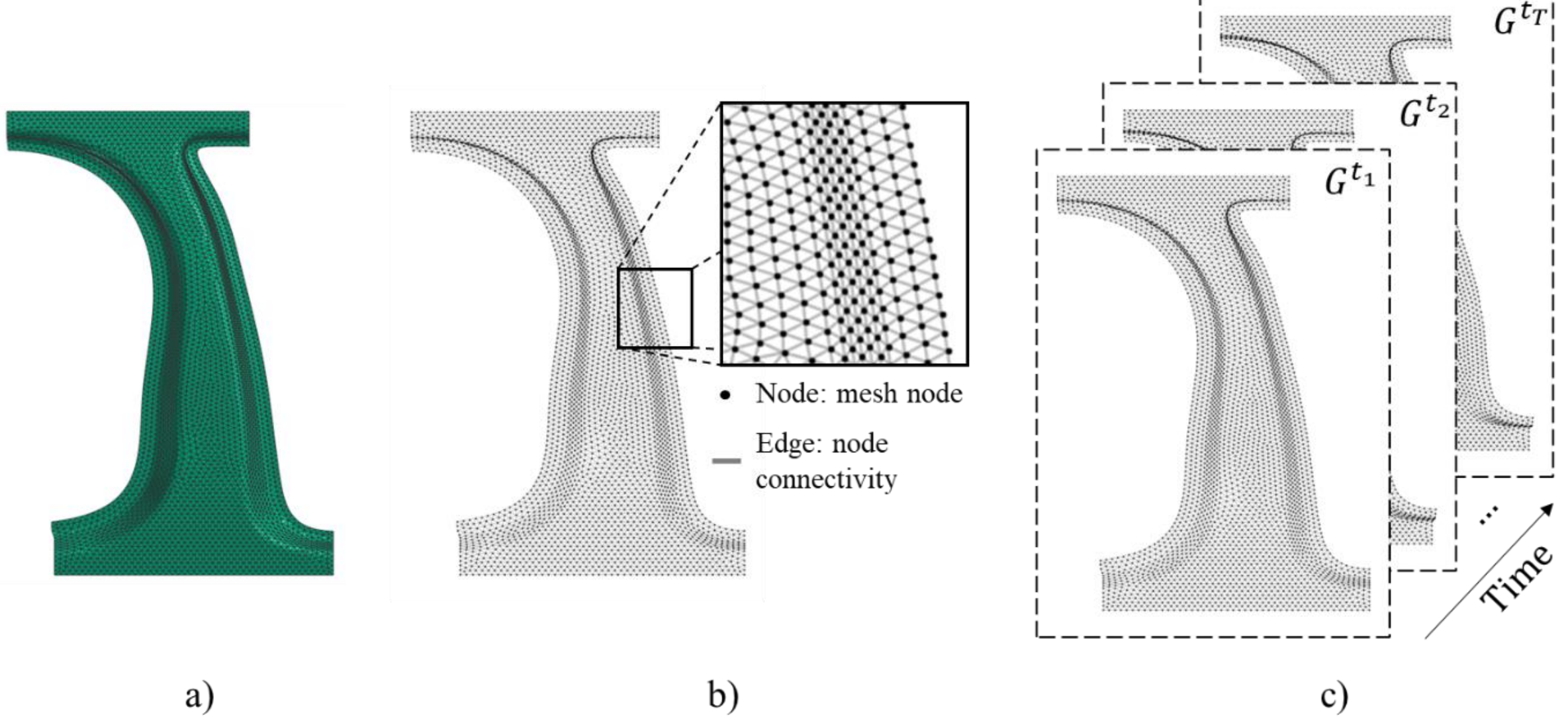

Figure 4: Illustrations of data representation for a B-pillar. a) FE mesh data of the B-pillar b) Graph representation of the B-pillar at a single time step c) The whole graph sequence consisting of $T$ graphs over time.

## 4.2 Multi-scale graph representation

One iteration of message passing propagates information from each node to its immediate neighbouring nodes. Solid mechanics simulations however often involve long-range dynamics due to stiff material, in which nodes can affect each other even if they are not in close proximity. Worse, crash simulations of vehicle components often require fine mesh configurations. This means that in order to propagate information over the graph, hundreds of iterations of message passing may be required. In order to reduce computational burden and improve model efficiency, multi-scale message passing is used, similar to MS-MGN [29]. Here, messages are being passed both on the fine-scale graph for local feature capturing, as well as on multiple levels of coarsened graphs, for faster information propagation. This message passing strategy is explained in detail in Section 4.3.

To solve solid mechanics-related problems, the coarsened graphs need to preserve the overall component structure and maintain the relative uniformity of the mesh. Therefore, coarsened graphs are created using FE meshes with larger element sizes. Figure 5 demonstrates the graph coarsening process of the B-pillar model. During each level of graph coarsening, the element size increases by a factor of 2, leading to a reduction of total number of nodes by a factor of 4. After 3 iterations of coarsening, the number of nodes reduces from around 3500 to 90, while preserving the overall shape and relatively uniform node distribution.

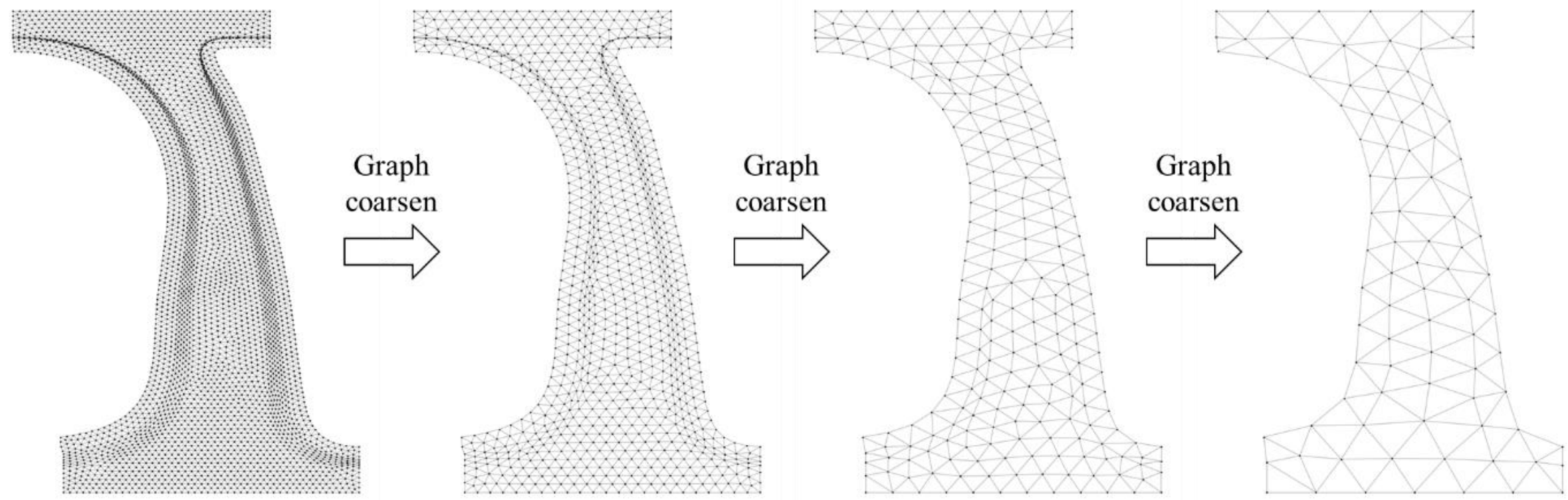

Figure 5: Demonstration of graph coarsening of the B-pillar component.

The coarsened graphs are generated from a benchmark sample, which has the initial geometrical design. This is the benchmark sample used for dataset generation via mesh morphing. As a result, the fine graphs from all samples are downsampled to the same set of coarsened graphs. This approach is chosen for two reasons: firstly, in component design optimisation tasks, despite the variation in design features across samples, all samples exhibit a similar overall structure; secondly, this ensures a fixed number of edges in the coarsened layers, which significantly improves both accuracy and efficiency during cross-graph message passing. This will be further explained in Section 4.3.4.

We use a graph U-Net architecture, with multiple layers of downsampling and upsampling. For passing information between graphs of different levels, edges are connected between the fine and coarse graphs. Figure 6 illustrates the edge connection between two different layers. Cross-graph edges are connected by linking each node in a fine graph to its k nearest neighbours in the corresponding coarse graph, based on spatial proximity. Figure 6 shows an example of such cross-graph edge connection of a single node when $k = 3$. This method ensures that every node from the fine layer remains connected to the coarse layer, avoiding any potential loss of information during downsampling and upsampling.

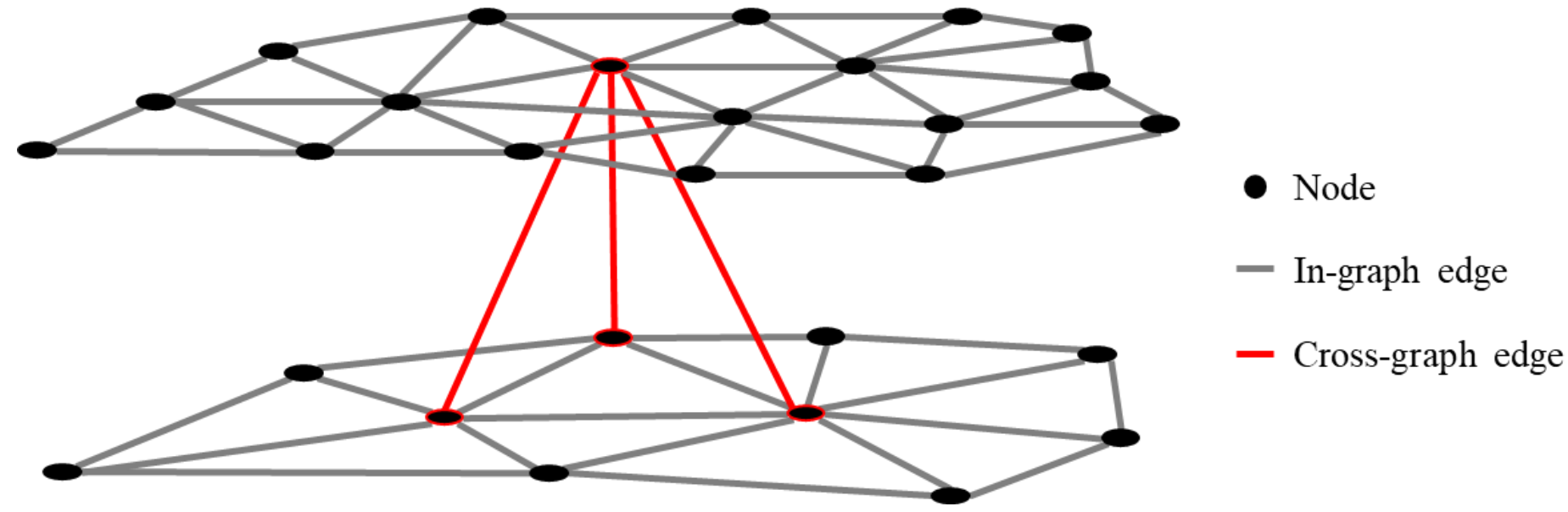

Figure 6: Illustration of the cross-graph edge connection for graph downsampling and upsampling processes, in this example, a fine graph node is connected to 3 nearest coarse graph nodes.

## 4.3 Recurrent Graph U-Net

In this section, the architecture of ReGUNet is detailly explained. An overview of the architecture is first provided, followed by details of each component of the model.

### 4.3.1 The overall architecture of ReGUNet

ReGUNet is designed for predictions of temporal graphs with time sequence structure. Previous studies [11, 17] have demonstrated the performance improvements achieved by recurrent architectures in temporal prediction tasks, therefore, adopting the idea, ReGUNet operates in a recurrent manner. Figure 7 a) illustrates the overall architecture of ReGUNet when predicting a graph sequence with $T$ time steps. The core of the model is the GUNet block, which performs prediction of the next time step based on the input from the current time step and the hidden state. An initial hidden state is input in the first iteration of prediction together with the first graph from the sequence. For each time step, the GUNet receives the hidden state from the previous time step $\mathbf{H}^{t_i}$ . It then outputs the updated hidden state $\mathbf{H}^{t_{i+1}}$ , which is directly fed into the next iteration, as well as the node features for the next-step input. Similar to Piers [17], the hidden state is updated with the edge features. Instead of GRU layers, we adopt a vanilla RNN layer for hidden state update for optimal computational efficiency, as the temporal sequences involved in this study are relatively short. This will be explained in more details in Section 4.3.3. As a multi-scale architecture, recurrent message passing is performed at both the finest and coarsest layers. The overall hidden state $\mathbf{H}$ includes two sets of hidden state matrices $\mathbf{H}_f$ and $\mathbf{H}_c$, for the finest and coarsest levels respectively. $T-1$ rounds of node feature predictions are made until the final time step is reached.

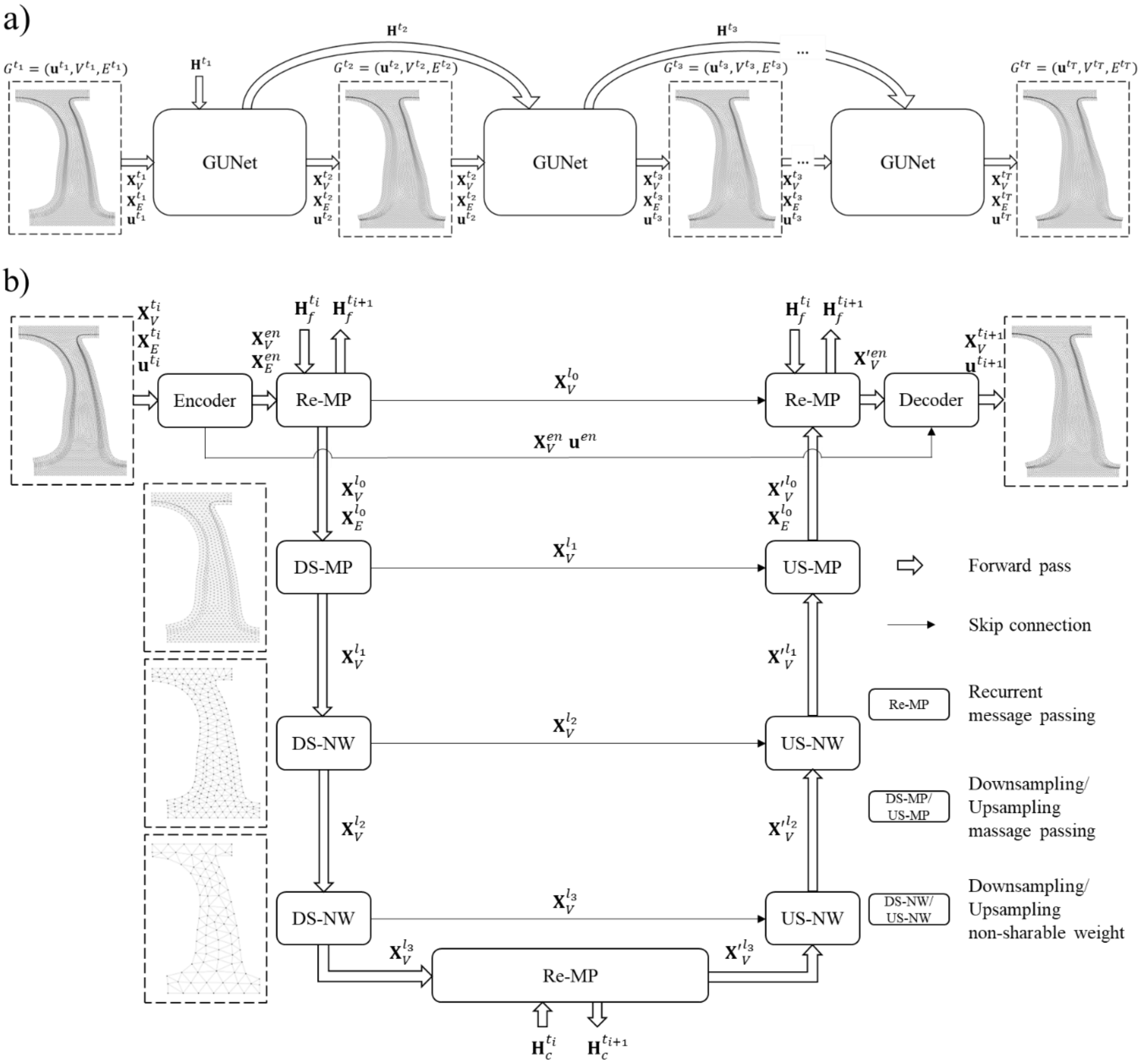

Figure 7: Illustrations of a) the overall recurrent architecture for prediction of a graph sequence, b) the architecture of the individual GUNet block from $t_i$ to $t_{i+1}$.

Figure 7 b) illustrates the internal architecture within a GUNet block. At time step $t_i$, the input graph with node and edge feature matrices $\mathbf{X}_V^{t_i}$ and $\mathbf{X}_E^{t_i}$, and optionally graph feature vector $\mathbf{u}^{t_i}$ first goes through an encoder which encodes the input features to the latent space, denoted as $\mathbf{X}_V^{en}$, $\mathbf{X}_E^{en}$ and $\mathbf{u}^{en}$. Together with the fine-layer-hidden-state $\mathbf{H}_f^{t_i}$ from the previous time step, the encoded node and edge features are then updated with a recurrent message passing (Re-MP) layer with $P_f$ message passing steps. The Re-MP layer returns the updated features for layer $l_0$, $\mathbf{X}_V^{l_0}$ and $\mathbf{X}_E^{l_0}$, as well as the hidden state $\mathbf{H}_f^{t_{i+1}}$which is directly fed into the next time step. This is followed by a number of downsampling layers propagating information to the most coarsened graph. The number of downsampling layers depends on the mesh density of the FE simulation data. Normally, a denser mesh requires more levels of graph downsampling to achieve efficient long-range message passing at the coarsest layer. In the example shown in Figure 7, three downsampling layers, including one message passing downsampling layer (DS-MP) and two non-shareable weight downsampling layers (DS-NW), are used. DS-NW assigns a unique weight to each cross-graph edge across all channels, enabling a more accurate message passing for nonlinear relationships. This leads to a limitation that all input samples must contain a fixed number of edges to be compatible with the weight matrix of the DS-NW. Note that edge features are not considered for the non-shareable weight layers as common coarsened graphs are used. To address this limitation, DS-MP, a more generalisable downsampling approach, is employed in the first downsampling layer. This approach leverages an MLP-based message passing mechanism, similar to the Re-MP layer, and is capable of processing graph structures of various connectivities with different numbers of nodes and edges. As a result, the model can accommodate geometries with previously unseen shapes and connectivities, enhancing its applicability to diverse and complex mesh configurations. For the Re-MP at the coarsest level, although in-graph edge feature matrix is used for message passing, it remains unchanged across different samples. As the edge feature matrix is not a variable, it is not included in Figure 7 b).

After downsampling, the features then go through another Re-MP layer with $P_c$ message passing steps. This layer performs message passing on the coarsened graph, while updating the coarse-layer-hidden-state matrix $\mathbf{H}_c$. As the coarse graph is much smaller, $P_c$ can be selected to be larger than $P_f$ without significant sacrifice in performance. The updated features are then upsampled with a set of upsampling layers that are symmetrical to the downsampling layers. Before each upsampling layer, a skip connection is employed to promote long-range information passing, forming a U-Net architecture. The final upsampled node feature $\mathbf{X'}_V^{l_0}$ are processed with $P_f$ steps of message passing by a Re-MP layer together with hidden state $\mathbf{H}_f$ and the previously updated edge feature $\mathbf{X}_E^{l_0}$**.** The outputs of this layer are then fed into the decoder, returning the final nodal prediction of the current time step, which is the incremental displacement fields of each node. This is also the nodal input for the next time step, $\mathbf{X}_V^{t_{i+1}}$, for autoregressive predictions. The edge feature of the next time step, $\mathbf{X}_E^{t_{i+1}}$, can be calculated based on the deformed shape of the component.

#### 4.3.2 Encoder and Decoder

Similar to GNS [33], the graph features are encoded into latent space with a channel number of $C_{en}$ by the encoder. Specifically, the node, edge, and graph features are encoded with 3 MLP sequences, each consisting of 4 fully connected layers followed by the ReLU activation function. Layer normalisation is employed to improve stabilisation during training. The encoder can be defined as:

$$\mathbf{X}_V^{en} = \mathbf{MLP}_{NodeEncoder}(\mathbf{X}_V), \qquad \mathbf{X}_E^{en} = \mathbf{MLP}_{EdgeEncoder}(\mathbf{X}_E), \qquad \mathbf{u}^{en} = \mathbf{MLP}_{GraphEncoder}(\mathbf{u}) \tag{2}$$

where $\mathbf{X}_V^{en}$, $\mathbf{X}_E^{en}$, and $\mathbf{u}^{en}$ are the encoded node, edge, and graph features respectively.

The decoder architecture depends on the desired output from the model. For node-level prediction only, the decoder consists of a single node-level MLP. When both node-level and graph-level outputs are required, an additional graph-level MLP branch is introduced. The output from the Re-MP layer, $\mathbf{X'}_V^{en}$, is not only fed into the node MLP for final nodal prediction, but is also mean pooled across all nodes to form a graph-level latent feature. This is then concatenated with the encoded graph feature, and processed by a graph MLP to obtain the graph-level output. Figure 8 shows the illustrations of both the encoder and decoder. The graph-level branch is optional and is activated only for tasks requiring graph-level responses.

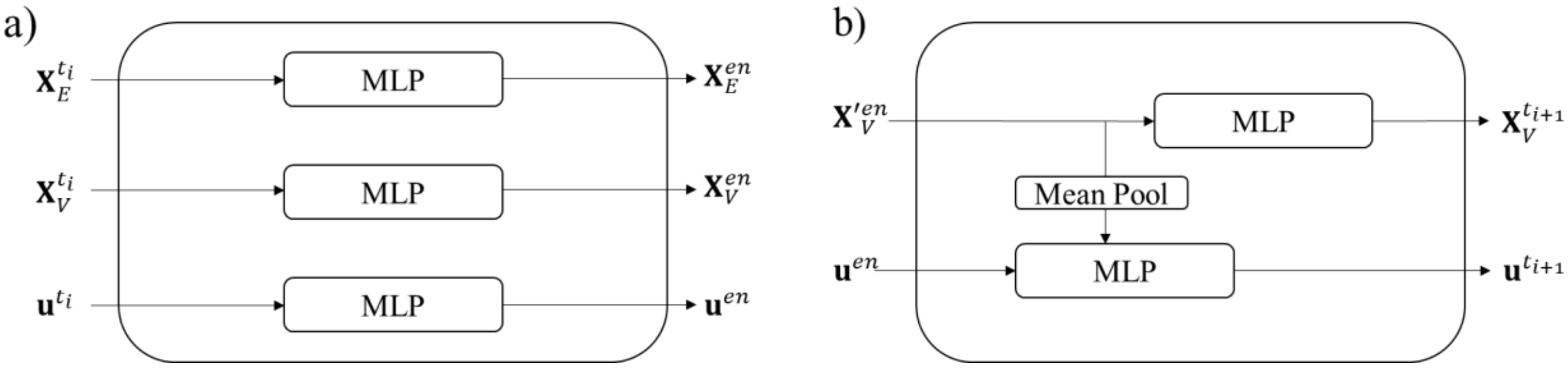

Figure 8: Illustrations of the architectures of the a) encoder, and b) decoder.

### 4.3.3 Recurrent message passing layer

As illustrated in Figure 7, the recurrent message passing (Re-MP) layer is employed in both the finest level $l_0$ and the coarsest level $l_3$ of the U-Net. The purpose of the first Re-MP layer is to aggregate information before the downsampling steps. The reason is that, geometric information is only encoded as edge features. In order to perform spatial computations at the coarse scale, this information has to first be integrated into the node feature vectors as the downsampling operator only passes node features to the next level. In addition, the Re-MP layer can aggregate and encode important non-smooth local features, which could otherwise get lost in the downsampling process. Because of the excessive number of edges in the fine graph, the number of message passing steps $P_f$ in this layer should be minimised to improve overall model efficiency.

The Re-MP layer in the coarsest level serves the purpose of a bottleneck in the architecture. This layer propagates and updates the concentrated node features through $P_c$ message passing steps to model the mapping between input and output features. Because of the reduced number of edges, message passing in this layer is highly efficient and a larger $P_c$ can be selected. Detailed hyper parameter tuning will be explained in Section 5.3. Another Re-MP layer is employed after upsampling to the finest level while referring to the initial features using a skip connection.

Each Re-MP layer consists of a recurrent edge block and a node block, updating the edge and node features iteratively. As the name suggests, the Re-MP layer introduces the recurrent operation across time steps by updating the hidden state. Two sets of hidden states are propagated through time: one at the finest level $\mathbf{H}_f$, and the other at the coarsest level of the U-Net $\mathbf{H}_c$. The hidden state matrices are input into the edge block together with the input edge features. Within each iteration of message passing, the features first pass through the edge block updating the edge information and the hidden state. The detailed algorithm for the edge block can be explained as:

$$\mathbf{X}'_E = \mathbf{MLP}_{Edge}\left(\mathbf{X}_{V_s}, \mathbf{X}_{V_r}, \mathbf{X}_E, \mathbf{H}\right), \qquad \mathbf{H}' = \mathbf{MLP}_{Hidden}\left(\mathbf{X}_{V_s}, \mathbf{X}_{V_r}, \mathbf{X}_E, \mathbf{H}\right), \tag{3}$$

where $\mathbf{X}'_E$ is the updated edge feature matrix, $\mathbf{X}_{V_s}$ and $\mathbf{X}_{V_r}$ are the node feature matrices of the sender and receiver nodes of all edges, $\mathbf{H} \in \mathbb{R}^{N \times p_e}$ is the hidden state, and $\mathbf{H}'$ is the updated hidden state. Note that the hidden state has the same dimension as the edge feature matrix, which is initialised to $\mathbf{0}$ at time $t_1$ and is updated iteratively at each time step to propagate local historical information. The updated edge information is then used in the node block for aggregating and updating the node features, which can be described as:

$$\mathbf{X}_V^{\text{agg}}[v_n] = \sum_{e_m \in E(v_n)} \mathbf{X}'_E\,[e_m]\,, \qquad \mathbf{X}'_V = \mathbf{MLP}_{Node}\left(\mathbf{X}_V, \mathbf{X}_V^{\text{agg}}\right), \tag{4}$$

where $E(v_n)$ is the set of edges connecting an arbitrary node $v_n$. Figure 9 illustrates the architecture of a Re-MP layer.

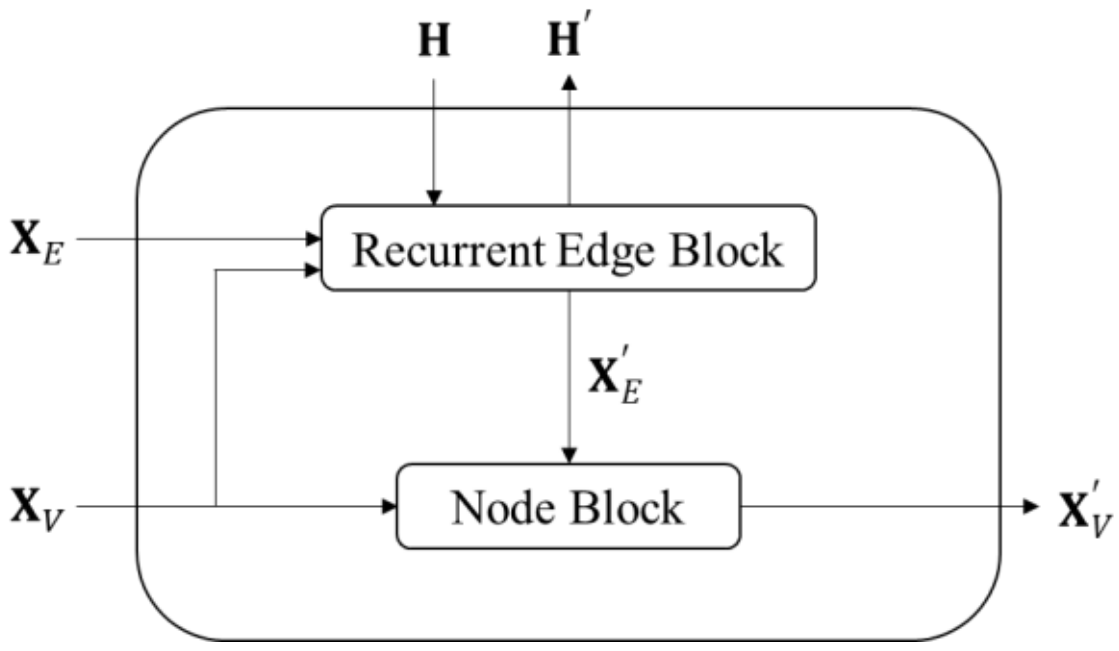

Figure 9: Illustration of the architecture of the recurrent message passing layer.

Multiple message passing steps within a graph is achieved by stacking a sequence of Re-MP layers. The number of message passing steps in each Re-MP layer is tuned as a hyperparameter, which is discussed in detail in Section 5.3. When the number of layers increases, the model may suffer from the vanishing gradient problem. To accommodate this, residual connection is added in each Re-MP layer by combining the input and output at each layer. This technique has been widely adopted in deep neural networks [46] and GNNs [31] as an effective way to mitigate vanishing gradients and enable deeper architectures.

#### 4.3.4 Downsampling and upsampling layers

As previously mentioned, the coarsened graphs are constructed using software generated FE meshes, which are shared across all samples within the dataset. However, the input graph of each sample has its own mesh configuration and therefore contain a different connectivity pattern and a different number of in-graph edges. As a result, the cross-graph connectivity between the input graph and the first coarsened graph is also different. To handle this, the architecture does not assume a fixed input graph topology. The downsampling from the input graph to the first coarsened graph is performed using the message passing downsampling layer (DS-MP), which can operate on arbitrary cross-graph connectivity. In contrast, the non-shareable weight down/upsampling layers (DS-NW / US-NW) are only used between the shared coarsened graphs, where the cross-graph topology is fixed across samples. Therefore, the fixed-topology requirement applies only to the shared coarse levels, not to the original input graph.

The DS-MP layer is similar to the Re-MP layer which performs message passing within a graph. A DS-MP layer propagates node features from the fine graph to the coarse graph through the cross-graph edges using their edge features consisting of 4 channels, the relative distances in the x, y, and z directions, and the Euclidean distance. Each DS-MP layer only performs one iteration of message passing, hence residual connection is not employed.

After the first downsampling layer, DS-NW layers are used for downsampling between the coarsened graphs. Inspired by the MAg layer [13], DS-NW first assigns a unique weight vector to each edge for feature update, making the weight matrix non-shareable. The assignment of a unique weight to each edge allows a more tailored aggregation of features, enabling the model to capture and adapt to more complex patterns within the data. The weight matrix can be denoted as $\mathbf{W} \in \mathbb{R}^{M \times C_{in} \times C_{out}}$, where $M$ is the number of edges, $C_{in}$ and $C_{out}$ are the input and output channel number to the layer. Similar to a standard U-Net, each downsampling layer increases the channel number by a factor of 2. DS-NW is especially suitable for feature aggregation with large channel numbers as it assigns channel-wise weights to each edge. DS-NW then aggregates the updated features to the neighbouring nodes. The algorithms can be expressed as:

$$\mathbf{x}'_{e_{v_s,v_r}} = \sum_{c \in C_{in}} \mathbf{x}_{v_s}^{(c)} \cdot \mathbf{w}_{e_{v_s,v_r}}^{(c)}, \qquad \mathbf{X}_{V_r}^{\text{agg}}[v_r] = LeakyReLU\left(\sum_{v_s \in N(v_r)} \mathbf{x}'_{e_{v_s,v_r}}\right), \tag{5}$$

where $\mathbf{x}'_{e_{v_s,v_r}} \in \mathbb{R}^{C_{out}}$ is the updated node features propagated through the edge connected from the sender node $v_s$ to the receiver node $v_r$, $\mathbf{x}_{v_s}^{(c)}$ is the $c$-th channel component of the node feature vector $\mathbf{x}_{v_s}$, $\mathbf{w}^{(c)}_{e_{v_s,v_r}} \in \mathbb{R}^{C_{out}}$ is the $c$-th channel component of the weight matrix $\mathbf{W}_{e_{v_s,v_r}}$ corresponding to edge $e_{v_s,v_r}$, $\mathbf{X}^{\text{agg}}_{V_r} \in \mathbb{R}^{N_{V_r} \times C_{out}}$ is the final aggregated node feature matrix of all the receiver nodes, $N(v_r)$ is the set of sender nodes that are connected to the receiver node $v_r$. The aggregated features are then fed through a LeakyReLU activation function to introduce nonlinearity.

The advantage of DS-NW compared to DS-MP is that the non-shareable weight matrix enables more accurate yet straightforward feature updates for multi-channel calculations. Each unique weight component is fully trainable, enhancing performance while maintaining high efficiency. However, this approach has poor generalisability as it requires the number of edges $M$ to be constant. A trained weight matrix cannot be directly used on other samples with different $M$. Therefore, DS-NW is only applicable for downsampling and upsampling between the commonly shared coarsened graphs. In contrast, DS-MP updates features based on MLPs, therefore can be applied to a variable number of edges, regardless of the graph size. The consideration of edge features before aggregation further enhances the generalisability of this approach. As a result, the message passing downsampling / upsampling layers are used at the finest level, and the non-shareable weight approach is applied to the coarser levels.

To conclude this section, we highlight the key innovations of ReGUNet relative to existing approaches. Unlike prior recurrence-based GNNs [17], ReGUNet integrates recurrence at both the finest and coarsest scales of a hierarchical U-Net architecture. Moreover, ReGUNet adopts a specialised non-sharable weight strategy for graph downsampling and upsampling, enabling more precise message passing to capture complex nonlinear relationships compared to existing multi-scale graph-based models [28, 29].

# 5 Results and discussion

ReGUNet is designed for temporal prediction, which can be trained either using a teacher forcing, autoregressive rollout, or scheduled sampling approach. For teacher forcing approach, at each time step, the model's predictions are based on the ground truth input from the graph sequence, rather than on its own previous prediction. This training technique helps guide the model by providing correct intermediate inputs, which aids in faster learning and reduces the risk of accumulating errors during training. Autoregressive rollout training approach involves recomputing the input graph based on the model's prediction of the previous time step. This approach is also used for model evaluation, where only the initial graph from every sequence is provided to the model. The model subsequently uses its own prediction of the initial graph to generate the second graph, and this process continues iteratively until the end of the entire sequence. This approach leads to error accumulation for later time steps which may affect training stability. As a compromise, the scheduled sampling training approach is also tested. Scheduled sampling interpolates between teacher forcing and autoregressive training by, at each time step, randomly choosing whether to feed the model the ground-truth graph or its own previous prediction, with the probability of using predictions gradually increasing over training. This reduces exposure bias and encourages the model to be robust to its own prediction errors during rollout. In this section, teacher forcing is used as the default training strategy unless explicitly stated otherwise, as it provides the most direct and controlled training setting for fair comparison across models and allows the comparison studies to focus on the target components themselves. The mean square error (MSE) is used as the loss function during training.

## 5.1 Baseline model comparisons

ReGUNet is compared with 5 Baseline models to evaluate its performance. In this section, Case study 1 is used and only node-level output is being predicted and evaluated. Baseline 1 is a CNN architecture, specifically, a Res-SE-U-net tailored for crashworthiness field prediction proposed by Li et al. [12]. The model takes in 4 channels of input images, consisting of a shape image and 3 displacement images. The shape image is the 2D projection of the deformed component shape from the z direction, and the displacement images are the x, y, and z incremental displacement fields from the previous time step. Baseline 2 is a variant architecture of MGN [27], which is an encoder-processor-decoder-based GN model. Baseline 3 is a variant of Piers [17], consisting of a GRU layer for edge recurrent function. Due the difference in the output prediction and the fairness of comparison, only the neural network architecture is adopted, where physics-informed loss function is not used in this study. Baseline 3 is a customised GN model that incorporates vanilla recurrence between time steps using the Re-MP layer, and the baseline is named as ReGNet. The key difference between ReGNet and ReGUNet is that ReGNet operates only on the original graph without any downsampling. Baseline 3 has a similar U-Net architecture as ReGUNet, but without recurrence between time steps. This model is named as GUNet. Baselines 1, 2, and 3 can be seen as external baseline models, where ReGNet and GUNet are variants of ReGUNet. The comparison between baselines 4 and 5 and ReGUNet serves as an ablation, which enables a clear visualisation of the performance improvement by introducing downsampling layers and hidden state propagation. All 6 models are trained for 1000 epochs using identical hyperparameter settings. The batch size, learning rate schedule, optimiser, and loss function are listed in detail in Table 1 to ensure fair and reproducible comparisons.

Table 1: Training hyperparameters.

| Hyperparameter | Value |
|---|---|
| Batch size | 2 |
| Learning rate (epochs 1-500) | 0.0004 |
| Learning rate (epochs 501-1000) | 0.0002 |
| Optimiser | Adam |
| Loss function | Mean squared error (MSE) |

The architectural configurations of the baseline models and ReGUNet are summarised in Table 2. For Res-SE-U-Net, 5 downsampling and upsampling layers are used to reach a channel number of 128 at the bottleneck. For MGN, Piers, and ReGNet, the encoder maps input features to a latent space with 128 channels similar to [17, 27]. This has been examined to be the optimal number of channels since further increases result in excessive GPU memory consumption. The hidden state dimension is kept the same with the edge feature dimension. 15 message-passing steps are used for the above mentioned GNN models. GUNet and ReGUNet use a hierarchical structure: the encoder begins with 32 channels, which double at each non-shareable weight downsampling layer, reaching 128 at the coarsest layer. GUNet and ReGUNet each uses 1 message-passing step at the finest layer and 15 at the coarsest layer.

Table 2: Model architecture summary.

| Model | Number of hidden channels | Recurrence | Downsampling strategy | Message passing steps |
|---|---|---|---|---|
| Res-SE-U-Net | 8, 16, 32, 64, 128 | None | Hierarchical (5 layers) | N/A |
| MGN | 128 | None | None | 15 |
| Piers | 128 | GRU | None | 15 |
| ReGNet | 128 | Vanilla RNN | None | 15 |
| GUNet | 32, 64, 128 | None | Hierarchical (3 layers) | 1 (fine), 15 (coarse) |
| ReGUNet | 32, 64, 128 | Vanilla RNN | Hierarchical (3 layers) | 1 (fine), 15 (coarse) |

All models are trained for 1000 epochs in a teacher forcing manner but are evaluated with autoregressive inference after training. The prediction accuracy is evaluated by calculating the mean point-wise Euclidean distance $D_t$ between ground truth and prediction during each time step, averaged over all nodes and all samples. For a given time step $t$, averaged across all $S$ samples:

$$D_t = \frac{1}{SN}\sum_{s=1}^{S}\sum_{n=1}^{N}\left\|\hat{\mathbf{x}}_{n,t}^{(s)} - \mathbf{x}_{n,t}^{(s)}\right\|_2, \tag{6}$$

where $\hat{\mathbf{x}}$ and $\mathbf{x}$ are the predicted and ground truth node features respectively. For the CNN baseline, image-to-node interpolation is performed to achieve fair comparison. Figure 10 a) shows the error accumulation of both training and validation data for all models over 12 time steps. Figure 10 b) shows the error distributions across the whole dataset at the final time step for all models.

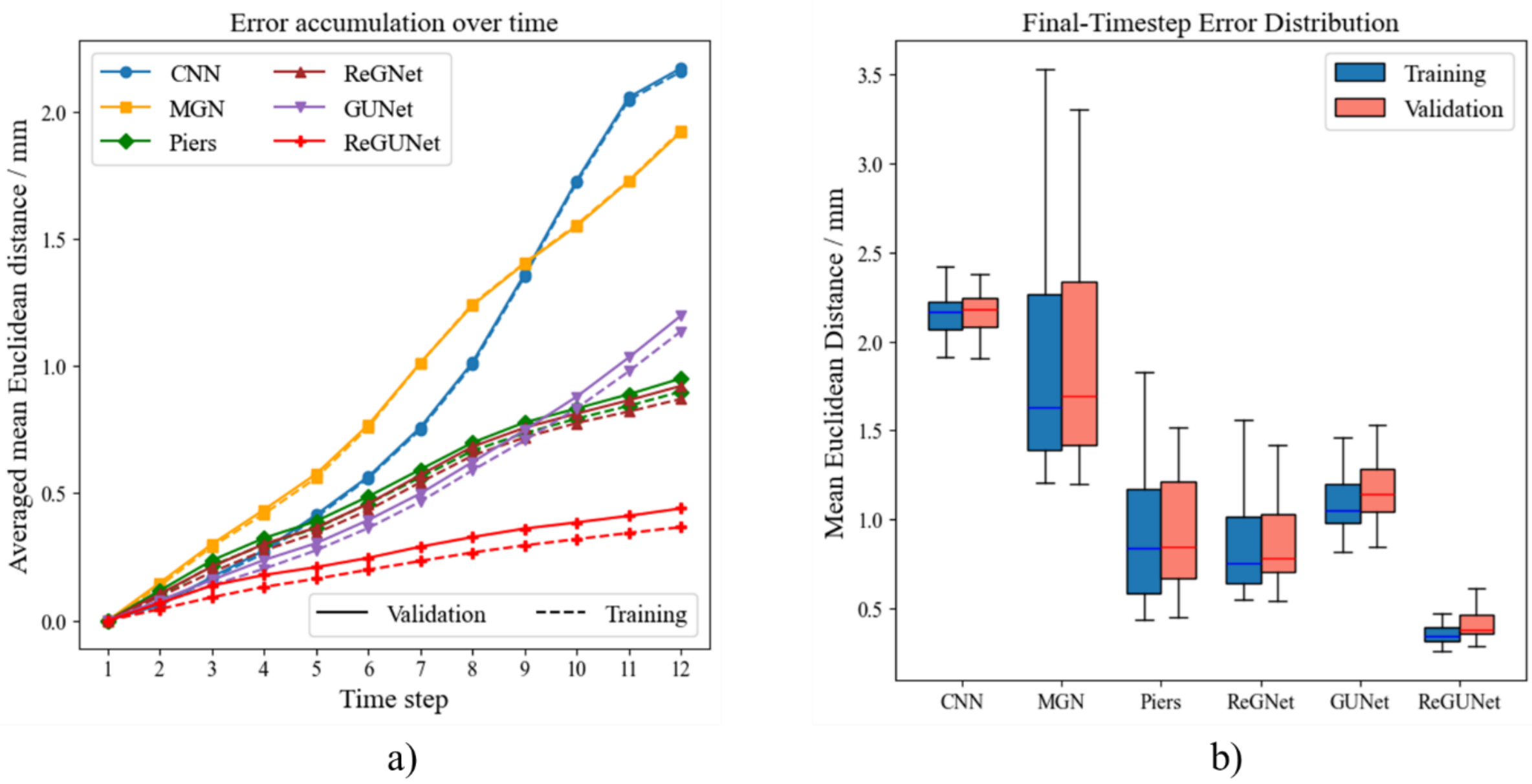

Figure 10: Comparisons between the baseline models in terms of a) error accumulation over 12 time steps, and b) final time-step error distributions, in terms of mean Euclidean distance.

As seen in Figure 10 a), all models exhibit increasing prediction error over time due to the compounding effect of autoregressive rollouts. However, clear performance differences emerge across architectures. the CNN baseline exhibits the largest growth in prediction error across time steps, indicating limited capability to preserve temporal consistency under autoregressive rollout. However, its error distribution in Figure 10 b) is relatively narrow compared to other baselines, suggesting consistently inaccurate but stable predictions. MGN demonstrates improved accuracy over CNN at later time steps, reflecting the advantage of graph-based spatio-temporal reasoning. Nevertheless, its error still accumulates rapidly as the rollout progresses, highlighting the limitations of purely local message passing in long-horizon prediction. Piers achieves prediction accuracy comparable to ReGNet and outperforms MGN in later time steps, but at the cost of significantly higher GPU memory consumption which will be discussed in detail later.

The comparison between MGN, ReGNet, GUNet, and ReGUNet form a structured ablation study to isolate the effects of temporal recurrence and graph downsampling. ReGNet extends MGN by introducing temporal recurrence, enabling the propagation of a hidden state across time steps. This design substantially reduces the rate of error accumulation, as evidenced by the slower growth of prediction error in Figure 10 a), particularly at later time steps. In contrast, GUNet incorporates hierarchical graph downsampling without temporal recurrence. While GUNet achieves lower errors than ReGNet at early time steps, its prediction error grows more rapidly over time and eventually surpasses that of ReGNet. This indicated greater temporal error propagation in the absence of recurrence. Nevertheless, the benefits of graph downsampling can be highlighted by comparisons between MGN and GUNet, as well as between ReGNet and ReGUNet. In both cases, the downsampled variants consistently achieve lower prediction errors and exhibit slower error growth across all time steps. This demonstrates that hierarchical downsampling significantly enhances spatial message passing by expanding the effective receptive field, enabling long-range interactions that cannot be efficiently captured by single-resolution message passing alone.

These trends are further verified by the final-timestep error distributions shown in Figure 10 b). The baseline models exhibit broader error distributions, whereas ReGUNet achieves both the lowest median error and the narrowest interquartile range, indicating superior accuracy and robustness. Notably, although ReGNet attains a lower median error than GUNet, its broader distribution reflects increased uncertainty due to the lack of spatial downsampling. Overall, these results demonstrate that temporal recurrence primarily mitigates long-term error accumulation, while graph downsampling improves spatial information propagation and prediction reliability.

Figure 11 compares the ground truth and predicted total z-displacement contours at the final time step for the 4 models in the ablation family - MGN, ReGNet, GUNet, and ReGUNet. One representative validation sample is selected for qualitative illustration. Two key observations can be drawn. First, for MGN and ReGNet, both of which operate solely on the fine graph without downsampling, the error fields are dominated by large, spatially coherent regions with the same error sign. This is because of insufficient message passing steps in these models. To propagate messages to physically distant nodes, significantly more message passing steps are required in fine graphs compared to coarse graphs due to higher node density with shorter edge connections. With 15 message passing steps, the distance information can travel in fine graphs is very limited, leading to less accurate prediction. Further increasing the number of message passing steps can result in a significant increase in GPU memory and inference time. In contrast, GUNet and ReGUNet, which employ graph downsampling, show improved contour agreement and reduced large-scale bias. Downsampling increases the physical distance covered per message-passing step on the coarsened graph, enabling information to propagate over a broader spatial extent using the same number of message-passing steps.

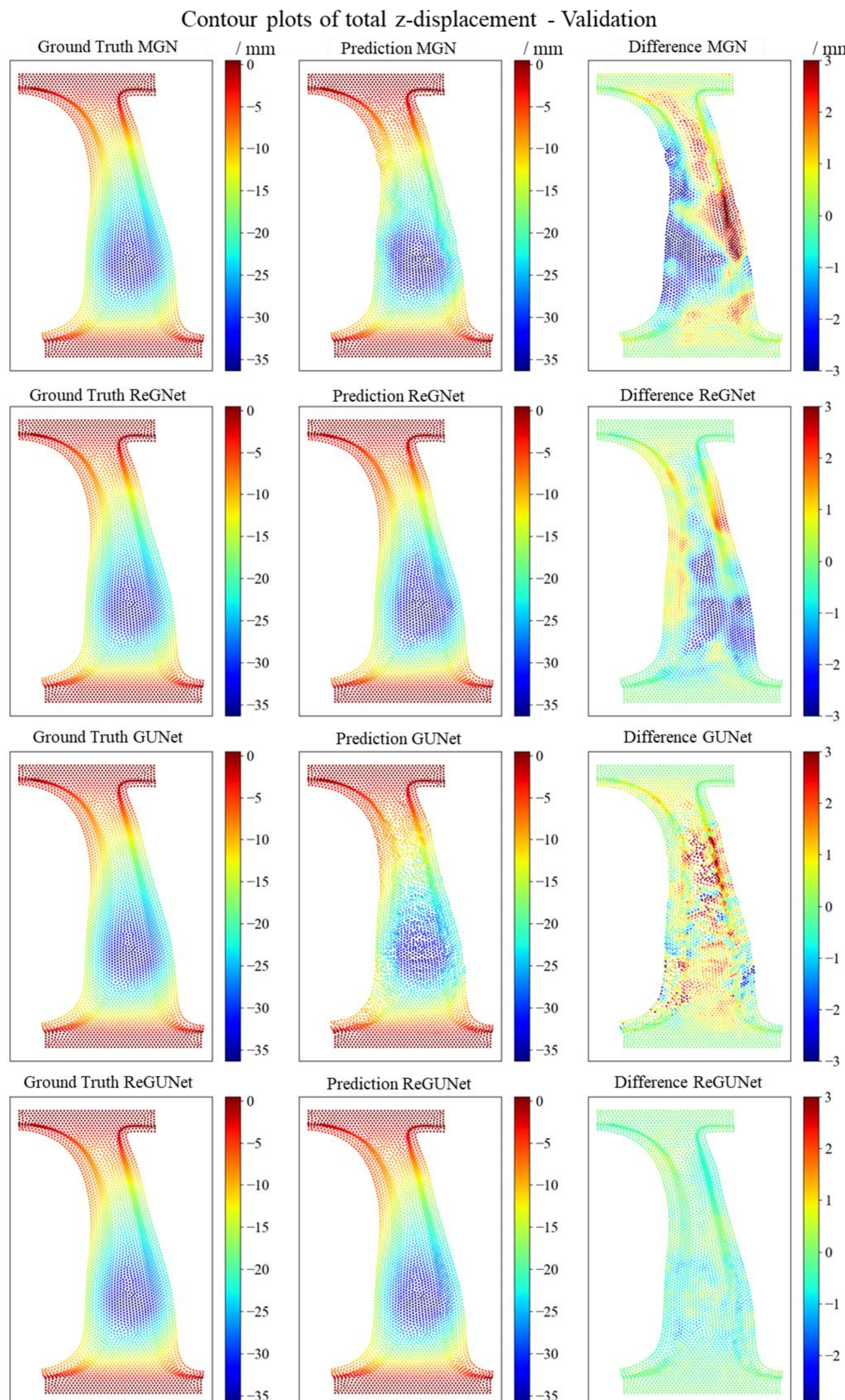

Figure 11: Comparisons between the four models in terms of total z-displacement contour plots of a validation sample in mm.

Secondly, one can also notice severe scatter-type errors that cause random waviness in the final predicted graphs for MGN and GUNet. As both models lack temporal recurrence, their predictions are more sensitive to locally inconsistent updates during autoregressive rollout, resulting in less smooth and less coherent graph structures. Unlike the large-scale error patterns caused by insufficient message-passing steps, these errors primarily affect local feature predictions. They arise from the accumulation of small errors over time, which become significant at later time steps. These small errors, occurring in multiple directions, lead to a wavy and scattered graph structure that becomes increasingly irregular as time progresses. The recurrence of hidden states, which contains historical information from previous time steps, helps mitigate the accumulation of errors over time. Therefore, models with recurrence are able to produce smoother graph structure predictions.

The training of all models is conducted using an Nvidia A100 GPU. Table 3 summarises the comparison among all models in terms of efficiency and accuracy. Reported means and 95% confidence intervals (CI) are computed using nonparametric bootstrap resampling (B=10,000 resamples). Among the external baselines, although having the lowest error range, the image-based Res-SE-U-Net yields the highest mean validation error (2.17 mm), indicating limited capability to preserve long-horizon structural evolution under autoregressive inference. MGN improves accuracy (1.92 mm) but still exhibits a wide error spread (0.68 mm), suggesting reduced robustness. Piers achieves competitive accuracy (0.95 mm) but requires the largest GPU memory footprint (32.1 GB) and the longest training time per epoch (1.46 min), highlighting a clear trade-off between performance and computational cost. This is due to the complexity brought by the GRU layers.

MGN, ReGNet, GUNet, and ReGUNet form a controlled ablation that isolates the effects of recurrence and graph downsampling. Comparing MGN to GUNet shows that downsampling significantly reduces GPU memory usage (15.9 GB to 8.7 GB) while also improving accuracy (1.92 mm to 1.20 mm), reflecting both improved scalability and better long-range spatial information propagation. Comparing MGN to ReGNet demonstrates that introducing recurrence improves long-horizon prediction accuracy (1.92 mm to 0.92 mm), despite a modest increase in computational cost (15.9 GB to 20.3 GB; 1.00 min to 1.24 min per epoch). Crucially, combining both mechanisms in ReGUNet yields the strongest overall performance. Relative to GUNet, adding recurrence reduces the mean validation error from 1.20 mm to 0.44 mm (≈63% reduction). Relative to ReGNet, incorporating downsampling reduces the error from 0.92 mm to 0.44 mm (≈52% reduction) while also reducing GPU memory consumption (20.3 GB to 9.5 GB). In addition to achieving the lowest mean validation error, ReGUNet exhibits a tight error distribution (standard deviation 0.14 mm) and the narrowest 95% CI [0.40, 0.48] mm, indicating reduced uncertainty and more reliable predictions. Finally, the pairwise p-values versus ReGUNet are far below conventional significance thresholds for all competing methods, providing strong statistical evidence that the observed improvements are not attributable to chance. Overall, ReGUNet offers the best balance of accuracy, robustness, and computational efficiency among all evaluated models.

Table 3: Overall comparisons between all the baseline models.

| | GPU memory / GB | Time per epoch / min | Mean validation error (bootstrap) / mm | Standard deviation of validation error / mm | 95% bootstrap CI of validation error / mm | P-values (vs. ReGUNet) |
|---|---|---|---|---|---|---|
| Res-SE-U-Net | **1.4** | **0.54** | 2.17 | **0.11** | 2.14 – 2.20 | $2.17\times10^{-50}$ |
| MGN | 15.9 | 1.00 | 1.92 | 0.68 | 1.74 – 2.11 | $5.28\times10^{-21}$ |
| Piers | 32.1 | 1.46 | 0.95 | 0.40 | 0.85 – 1.06 | $1.32\times10^{-15}$ |
| ReGNet | 20.3 | 1.24 | 0.92 | 0.39 | 0.82 – 1.04 | $1.54\times10^{-13}$ |
| GUNet | 8.7 | 1.14 | 1.20 | 0.27 | 1.12 – 1.27 | $2.06\times10^{-25}$ |
| ReGUNet | 9.5 | 1.25 | **0.44** | 0.14 | **0.40 – 0.48** | N/A |

## 5.2 Effect of NW-based cross-graph message passing

In this section, different approaches of cross-graph message passing, i.e., information transfer between resolutions, are evaluated with Case study 1. To isolate the effect of cross-graph message passing, 4 non-recurrent variants are evaluated so that performance differences are attributed primarily to the downsampling strategy rather than recurrence. The first variant, No DS, corresponds to the MGN-style model operating purely on the fine graph. The second, Variable DS-MP, uses a hierarchical representation where each downsampled graph is generated by independent software-based remeshing per sample similar to [29]. The remeshing is done using Hypermesh, making sure that the coarse graphs preserve the exact geometry of the fine graph but differ in connectivity. In this case, only DS-MP can be used for cross-graph message passing at all inter-level links because of varying number of edges. The third, Fixed DS-MP, uses a single benchmark hierarchy shared across all samples, again employing DS-MP for cross-graph message passing. Finally, Fixed DS-MP/NW corresponds to the GUNet-style setting in the previous section, where cross-graph transfer is handled by a combination of DS-MP and DS-NW on the shared benchmark hierarchy.

Figure 12 a) shows the error accumulation curves under autoregressive inference, and Figure 12 b) reports the final-timestep error distributions. First, introducing a multi-resolution hierarchy is clearly beneficial. All downsampled variants outperform No DS across all time steps, indicating that downsampling improves long-range information propagation and improves prediction accuracy. This aligns with the intuition that, on fine meshes, a moderate number of message passing steps covers only a limited physical range, where coarsening effectively increases the receptive field.

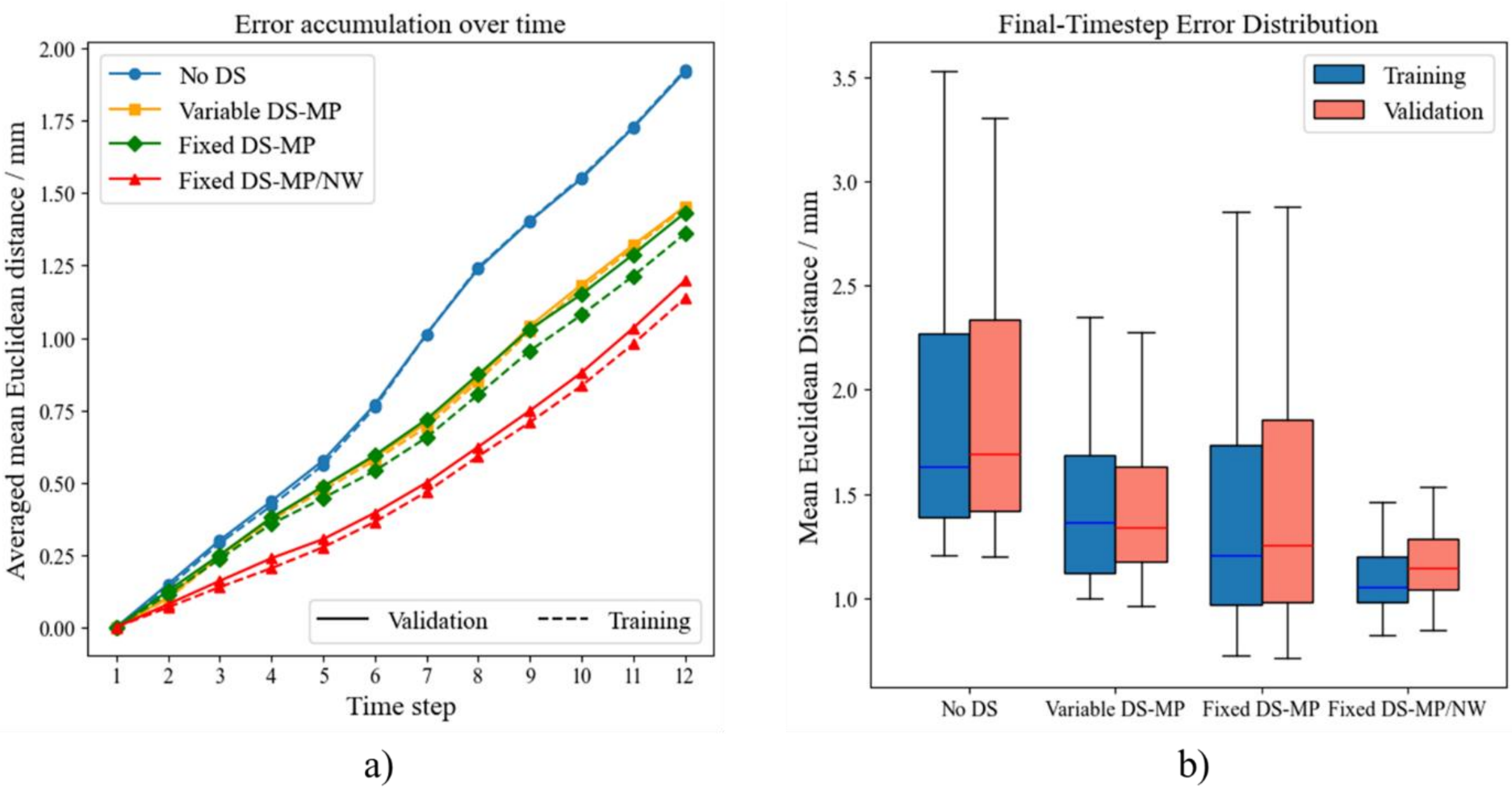

Figure 12: Comparisons between the different cross-graph downsampling approaches in terms of a) error accumulation over 12 time steps, and b) final time-step error distributions, in terms of mean Euclidean distance.

Second, the choice between variable and fixed downsampled graphs has only a minor effect on the mean error trajectory for this case study. Variable DS–MP and Fixed DS–MP exhibit very similar average error growth over time. However, the Fixed DS–MP variant shows a noticeably wider error spread at the final time step as shown in Figure 12 b). This suggests reduced robustness across samples when a single shared hierarchy is imposed. In addition, the temporal curves show a small but visible generalisation gap for the fixed-hierarchy setting, indicating mild overfitting. Nevertheless, the gap remains limited, implying that using a shared benchmark hierarchy is still a reasonable and scalable design choice. However, a fixed connectivity for coarsened hierarchy is required to employ DS-NW, as it relies on specific edge weight assignment. One can observe that replacing DS-MP with DS-NW for coarsened levels yields the largest improvement. The Fixed DS-MP/NW model consistently achieves the lowest accumulated error over all time steps, and simultaneously produces a tighter final-timestep compared to both DS–MP variants. This indicates that the NW mechanism not only reduces the average error, but also improves reliability by suppressing sample-to-sample variability. Overall, these results suggest that while hierarchical downsampling is the key enabler for long-range spatial reasoning, DS-NW is critical for accurate and stable cross-graph information transfer, leading to the lowest mean error. Note that the current benchmark evaluates samples with moderate shape variability, where the geometries remain broadly comparable. How these trends change under larger geometric variations will be explored in future work.

## 5.3 Hyperparameter tuning

To achieve the best performance with ReGUNet, it is essential to determine the optimal combination of hyperparameters, including the numbers of message passing layers $P_f$ and $P_c$, the number of cross-graph edges per node between finer and coarser layer $k$, and the channel numbers. The model is trained with different combinations of hyperparameters with Nvidia T4 GPU with 16 GB RAM provided by Google Colab. The random seed used in this section is 36, the effect of changing random seeds by run-to-run variations, presented in Appendix B.

Firstly, the effect of channel number on model performance is studied. The channel number output from the encoder is varied between 16, 32, 64. After 2 non-shareable weight downsampling layers, the channel number at the coarsest level reaches 64, 128, 256. Table 4 summarises the model's efficiency and accuracy with different channel numbers. The message passing steps are kept constant at $P_f = 1$ and $P_c = 15$; the number of cross-graph edges per node is kept at $k = 6$. When the channel numbers increase to 64 - 256, the GPU memory consumption exceeds the limit of 16 GB, making the model inexecutable. Comparing 16 - 64 and 32 - 128, the latter leads to a lower validation error despite an insignificant trade-off in efficiency in terms of both GPU memory and training time. Therefore, 32 - 128 is selected for further calibrations.

Table 4: The effect of channel number on model performance.

| Channel numbers | GPU memory / GB | Time per epoch / min | Mean validation error / mm |
|---|---|---|---|
| 16 - 64 | 6.7 | 1.02 | 0.66 |
| 32 - 128 | 9.5 | 1.24 | 0.45 |
| 64 - 256 | >16 | N/A | N/A |

Next, four values of $k$ between 3 and 12 are tested to determine the optimal number of cross-graph edges. With more cross-graph edges between two graphs, the receiver nodes receive information from a larger number of sender nodes. As before, $P_f$ and $P_c$ are kept at 1 and 15 respectively. Table 5 and Figure 13 show the effect of $k$ on model performance in terms of validation accuracy and efficiency. One can observe that when $k$ exceeds 6, both training time and memory consumption increase rapidly, however, the reduction in validation error gradually diminishes. This suggests that while increasing $k$ initially improves the model's accuracy, beyond a certain point, the computational costs outweigh the performance gains. As a result, $k = 6$ is chosen for further calibration.

Table 5: The effect of number of cross-graph edges per sender node on model performance.

| k | GPU memory / GB | Time per epoch / min | Mean validation error / mm |
|---|---|---|---|
| 3 | 7.6 | 1.39 | 0.57 |
| 6 | 8.5 | 1.40 | 0.45 |
| 9 | 9.2 | 1.48 | 0.42 |
| 12 | 9.5 | 1.63 | 0.40 |

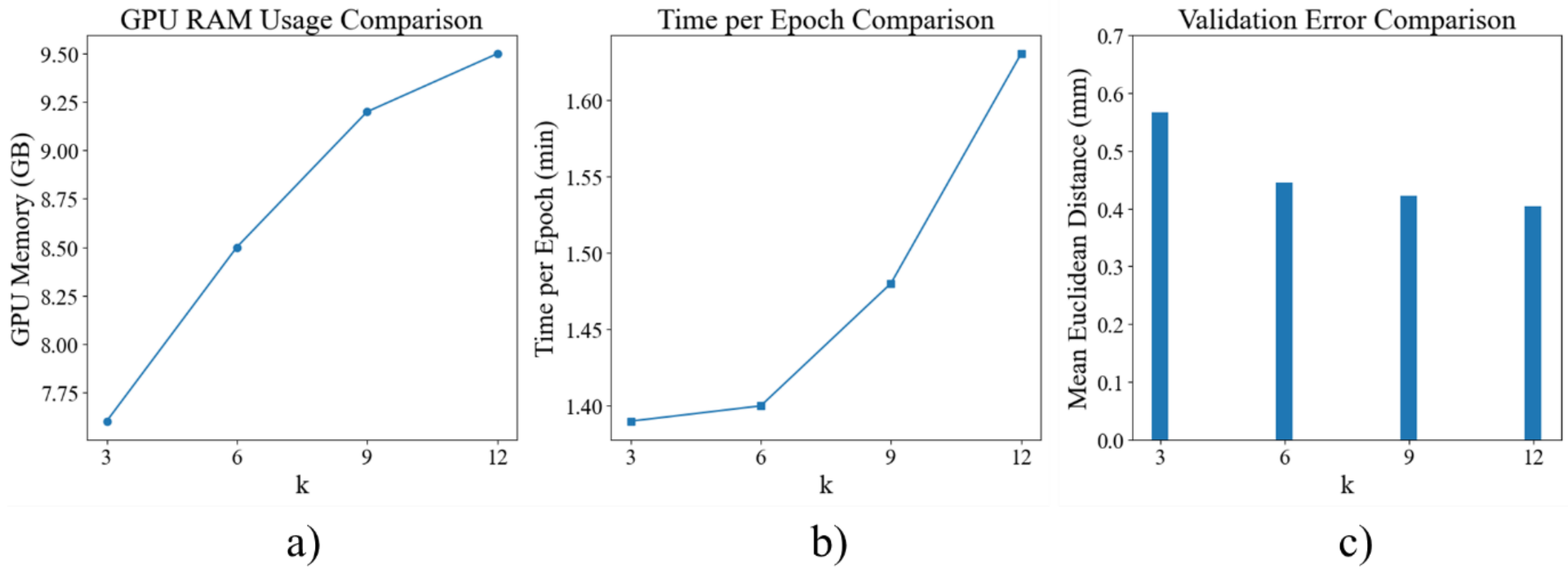

Figure 13: Illustrations of the effect of number of cross-graph edges per sender node on model performance, in terms of a) GPU memory consumption, b) training time, and c) validation error in terms of mean Euclidean distance.

Lastly, the optimal numbers of message passing steps $P_f$ and $P_c$ are determined. Eight combinations of message passing steps are evaluated, where $P_c$ taking values of 5, 10, 15, and 20, $P_f$ is set to 1 and 2. A larger $P_f$ will result in memory consumption exceeding the available limit. This is because message passing on fine graphs is more computationally expensive. Table 6 and Figure 14 summarise the relationship between message passing steps and model performance. One can observe that when $P_f = 2$, the validation error decreases, with a slight increase in training time. For $P_c$, 10 message passing steps shows the best prediction accuracy with the lowest GPU memory consumption. Meanwhile, the training time is significantly lower than that when $P_c$ is 15 and 20. The reason why more message passing steps leads to a higher error may be due to the oversmoothing problem. This occurs when, after a certain number of layers, the node features become increasingly similar, making it difficult for the model to distinguish between different nodes. As message passing continues, the node features converge, homogenising the node representations and reducing the model's ability to capture local differences. One can also observe that further increasing the number of message passing layers to 20 results in a slight recovery in validation performance. This may be attributed to the model's increased capacity to capture more complex or long-range dependencies within the graph structure due to the extra message passing steps. As a result, 2 and 10 message passing steps for the fine and coarse levels are selected as the final optimised parameters.

Table 6: The effect of numbers of message passing steps on model performance.

| | GPU memory / GB | | Time per epoch / min | | Mean validation error / mm | |
|---|---|---|---|---|---|---|
| | $P_f = 1$ | $P_f = 2$ | $P_f = 1$ | $P_f = 2$ | $P_f = 1$ | $P_f = 2$ |
| $P_c = 5$ | 7.3 | 9.2 | 0.76 | 0.98 | 0.40 | 0.36 |
| $P_c = 10$ | 7.0 | 9.6 | 1.10 | 1.26 | 0.38 | 0.30 |
| $P_c = 15$ | 8.5 | 10.1 | 1.40 | 1.63 | 0.45 | 0.36 |
| $P_c = 20$ | 8.7 | 10.1 | 1.73 | 1.86 | 0.44 | 0.34 |

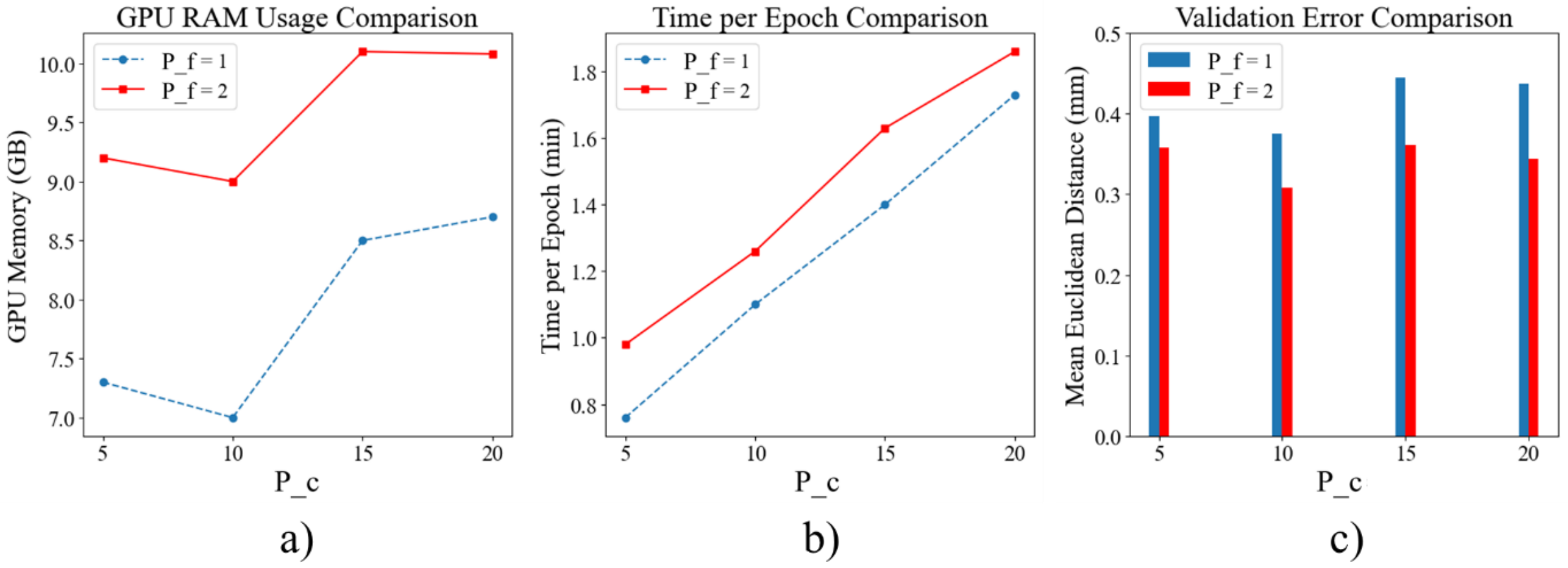

Figure 14: Illustrations of the effect of numbers of message passing steps on model performance, in terms of a) GPU memory consumption, b) training time, and c) validation error in terms of mean Euclidean distance.

Finally, the model with 6 cross-graph edges per sender node, channel numbers of 32 - 128, and message passing steps of 2 and 10 in the fine and coarse levels, is selected as the final tuned model. Figure 15 first compares the error accumulation of the tuned and untuned model for the training set and validation set. In comparison to the untuned model, the error accumulation is significantly reduced, indicating that the tuning process has effectively enhanced the model's performance by minimising the propagation of errors, resulting in more accurate and consistent predictions. In addition, the tuned model is further evaluated with the test set. The averaged error accumulation across the test set is also plotted in Figure 15. One can observe that compared to the validation set, there is a very slight increase in terms of mean Euclidean distance at later time steps, reaching a final test error of 0.32 mm. This demonstrates the model's ability to generalise and make accurate predictions on previously unseen samples.

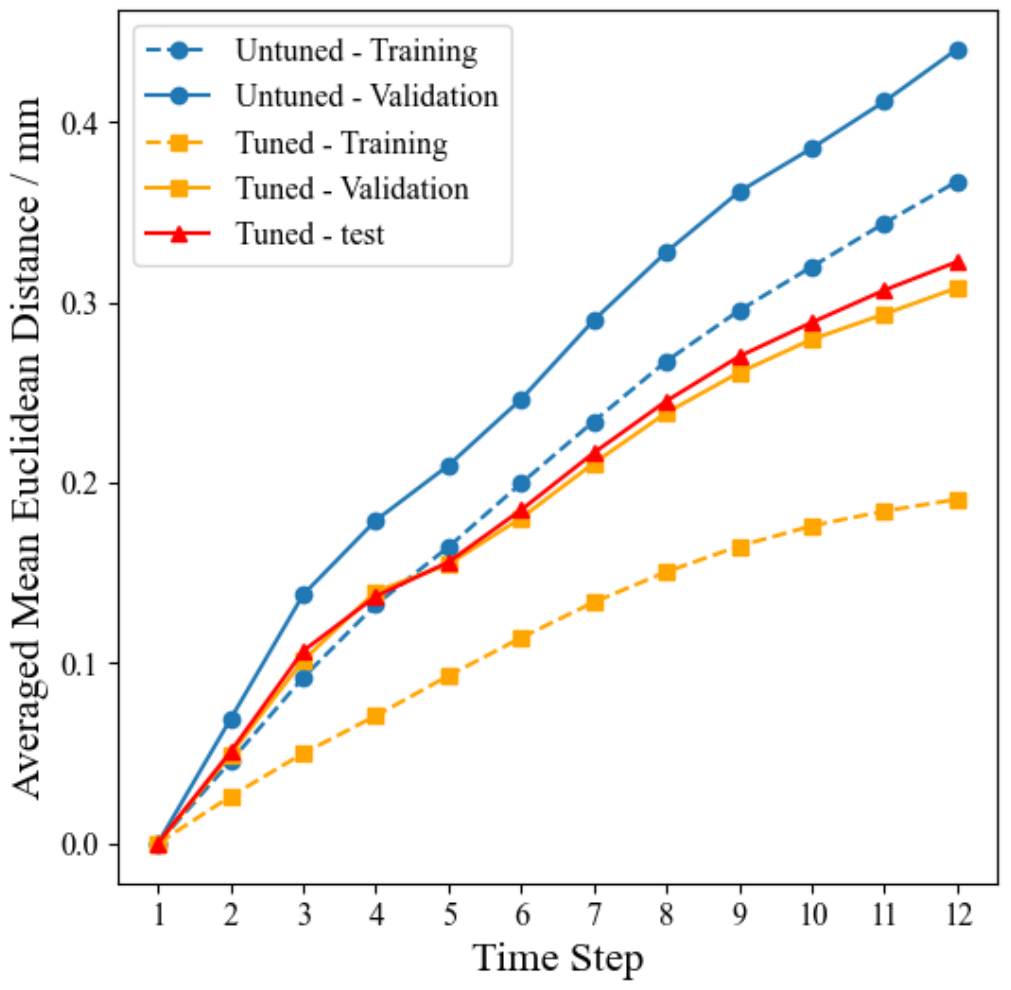

Figure 15: Error accumulation of the ReGUNet after hyperparameter tuning in terms of mean Euclidean distance.

Figure 16 shows representative samples in the test set showing the prediction of the total z-displacement field of the B-pillar. One can observe good agreement between the ground truth and prediction. Also, the model is able to capture the change in resultant deformation fields given the variation in component design. One of the key metrics used to evaluate crashworthiness performance is the maximum B-pillar intrusion, which is defined as the maximum z-displacement of the B-pillar during a crash test. This is the maximum z-displacement value extracted from the final time step for each sample. The percentage error of each sample is calculated as $\frac{|\hat{z}_{max}-z_{max}|}{|z_{max}|} \times 100\%$, where $\hat{z}_{max}$ and $z_{max}$ denote the predicted and ground truth maximum z-displacement values, respectively. The mean percentage error of the B-pillar intrusion across all the test set is 0.74%. To quantify run-to-run variation induced by stochastic training effects, the model is trained and evaluated using different random seeds. Results show that the performance improvement is reproducible across multiple training runs, and is not driven by a single favourable seed. More details can be found in Appendix B. The full prediction of the dynamic behaviour of a representative sample in the test set is provided in Appendix C.

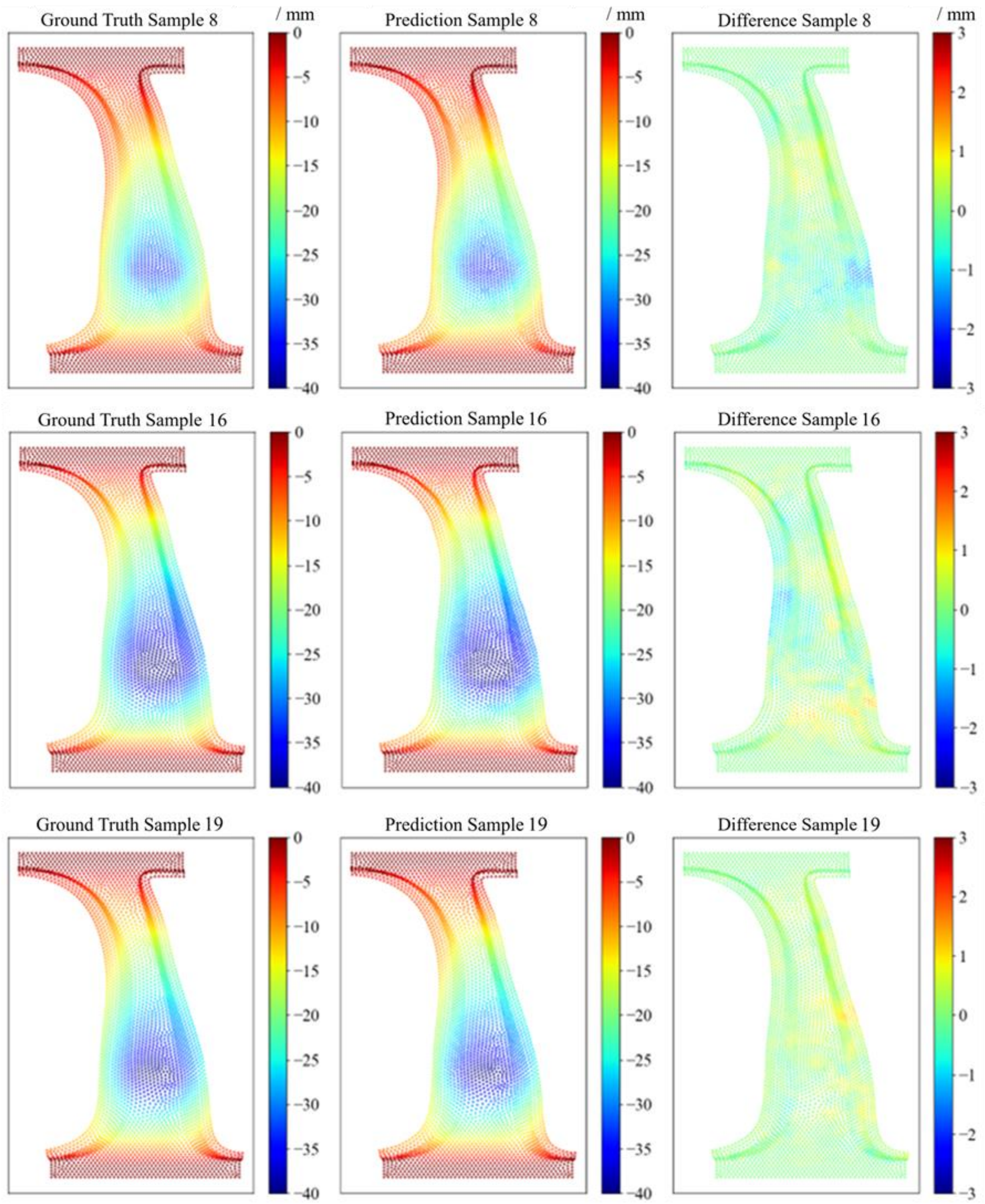

Figure 16: Differences between ground truth and prediction of total z-displacement of representative test samples in mm.

## 5.4 Effect of training strategy

This study considers three training strategies for learning temporal crash responses, which are teacher forcing, scheduled sampling [47], and rollout (fully autoregressive) training. All trained models are evaluated under the same autoregressive inference setting. Teacher forcing trains the model by conditioning each prediction on the ground-truth state from the previous time step. While this often leads to fast and stable optimisation, it introduces a mismatch between training and inference. This is because at test time, the model must condition on its own predictions instead of the ground truth. This "exposure bias" is reflected in Figure 17 a), where the teacher-forcing model exhibits the steepest increase in deformation error over time, and in Figure 17 b), where the final-step validation error distribution is higher and broader than the other strategies. This shows that teacher forcing learns accurate one-step mappings under ideal inputs, but is less robust when the input sequence becomes progressively imperfect during closed-loop rollout.

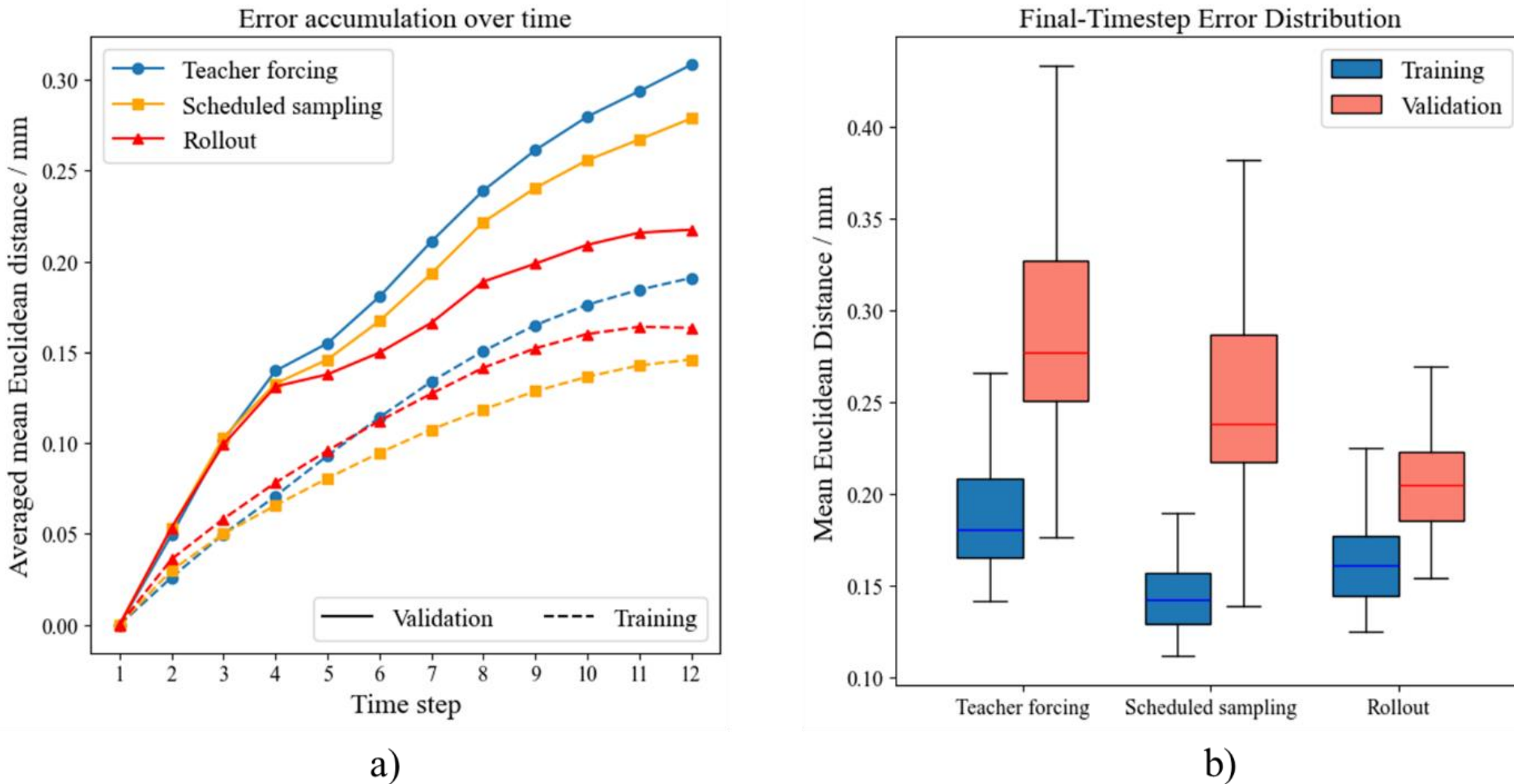

Figure 17: Comparisons between the different training strategies in terms of

a) error accumulation over 12 time steps, and b) final time-step error distributions, in terms of mean Euclidean distance.

Scheduled sampling [47] combines teacher forcing and autoregressive inputs during training. Specifically, the next-step input is chosen as either the ground-truth state or the model prediction, controlled by a teacher-forcing ratio (tf-ratio). In this implementation, the teacher-forcing ratio is linearly decayed from 1 to 0 across epochs, so the model transitions gradually from fully supervised conditioning to fully autoregressive conditioning. This approach reduces exposure bias without making early training unstable. As shown in Figure 17 a), scheduled sampling consistently reduces the slope of error accumulation compared with pure teacher forcing, and Figure 17 b) shows a lower mean final-step error with a comparably tight uncertainty range, indicating improved stability.

Finally, rollout training uses tf-ratio = 0 throughout training, meaning that the model always feeds back its own predictions when constructing the next input. This strategy most closely matches the inference regime and therefore directly optimises the model under the same conditions as inference, using its own predictions as inputs. As seen in Figure 17 a), although having slightly higher training error, rollout training produces the slowest error growth and the lowest final-step deformation error for the validation set. The statistics in Figure 17 b) also indicate the most favourable distribution, where rollout training achieves the lowest mean and the narrowest spread. Conventionally, rollout training can be difficult in long-horizon problems at early epochs, since early prediction noise may destabilise optimisation. However, in this work the horizon is relatively short and the error accumulation does not grow to a level that causes training divergence. Therefore, the advantage of eliminating the train–test mismatch dominates, making rollout training the most effective approach for improving long-term consistency. This can also be demonstrated by the fact that rollout training leads to the smallest generalisation gap between training and validation accuracy.

Overall, these results suggest that when the prediction horizon is modest and the underlying dynamics are learnable without severe compounding instability, matching the training regime to the inference regime is beneficial. Scheduled sampling provides a robust intermediate solution, while full rollout training yields the best performance by explicitly training the model to remain stable under its own recursive predictions.

### 5.5 Extending to wider design space and multi-task outputs

In this section, ReGUNet is trained and evaluated on Case Study 2, which spans a wider design space and requires simultaneous node-level and graph-level prediction for a more comprehensive crashworthiness assessment. At each time step, the model outputs (i) three channels of nodal displacement and (ii) two scalar KPIs, namely the B-pillar's internal energy (used here as an estimation for energy absorption during the crash event) and the contact force between the impactor and the component. Different from the displacement fields, the scalar KPIs are learned in normalised absolute form (rather than incremental form) to reduce error accumulation over autoregressive rollouts and to stabilise the sequence-level trend prediction. The model is trained with teacher forcing training strategy, and the hyperparameters are identical as described in Section 5.3.

Figure 18 summarises the prediction performance of the tuned model. Figure 18 a) reports deformation accuracy using the averaged mean Euclidean distance between predicted and ground-truth nodal coordinates under autoregressive inference. As expected, the deformation error increases with time due to error propagation, but remains well controlled, reaching below 0.5 mm at the final time step. Figure 18 b) and c) show scatter plots of predicted (Pred) versus ground-truth (GT) scalar KPIs at the final step with the coefficient of determination ($R^2$), together with the $y = x$ reference line. The model achieves strong agreement for internal energy (Train $R^2 = 0.942$, Test $R^2 = 0.892$) and contact force (Train $R^2 = 0.940$, Test $R^2 = 0.850$), indicating that the learned global response trends generalise well despite the enlarged design space.

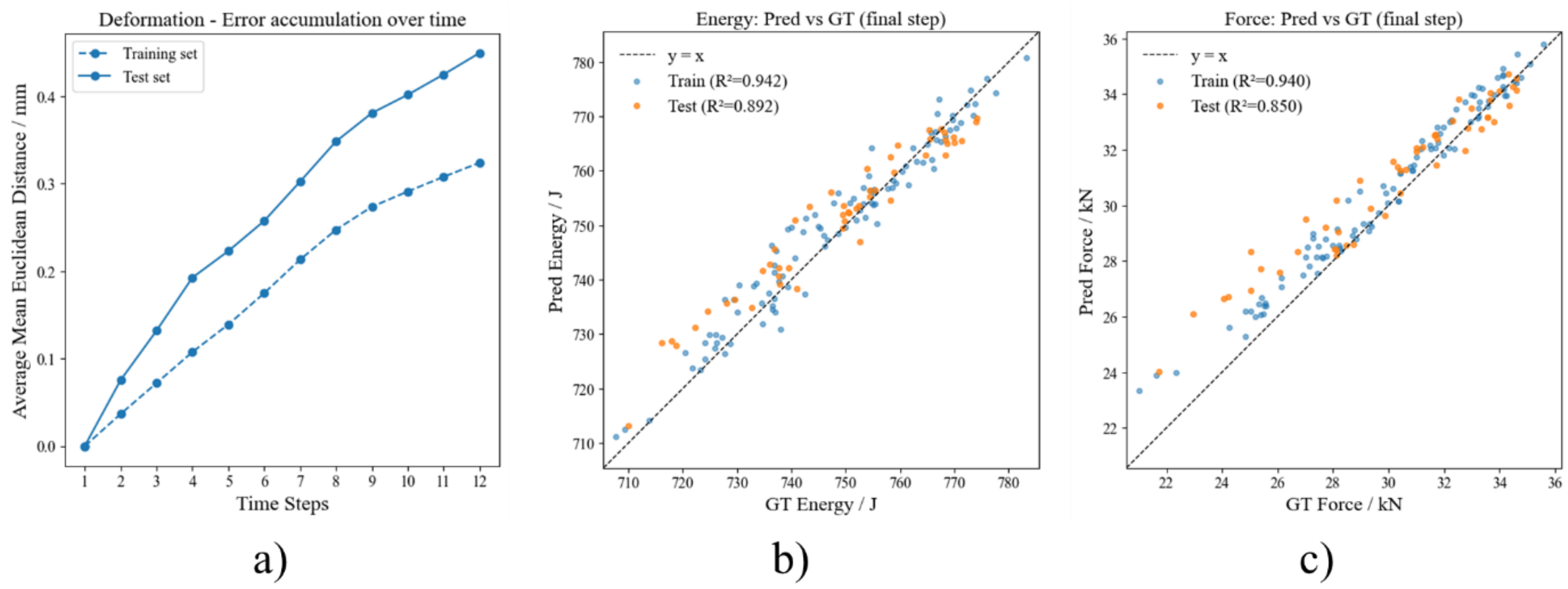

Figure 18: Illustration of the prediction results: a) the mean deformation error over time, prediction vs. ground truth scatter plots for b) internal energy, and c) contact force.

To further interpret these scalar results beyond aggregated scatter statistics, Figure 19 presents time-history comparisons for three representative test samples. For both energy and force, the predicted trajectories closely track the ground truth across the full sequence, capturing the overall rise and the subsequent turning behaviour near the late stages. Minor deviations are mainly observed in the force prediction due to greater variations of the force-time curve compared to the energy-time curve. This is consistent with the lower test $R^2$ for force in Figure 18 c). Overall, these examples confirm that the model not only matches the final-step values, but also preserves the temporal evolution of the engineering responses. Appendix D shows the detailed prediction accuracy during intermediate time steps of the two scalar outputs in terms of scatter plots and $R^2$.

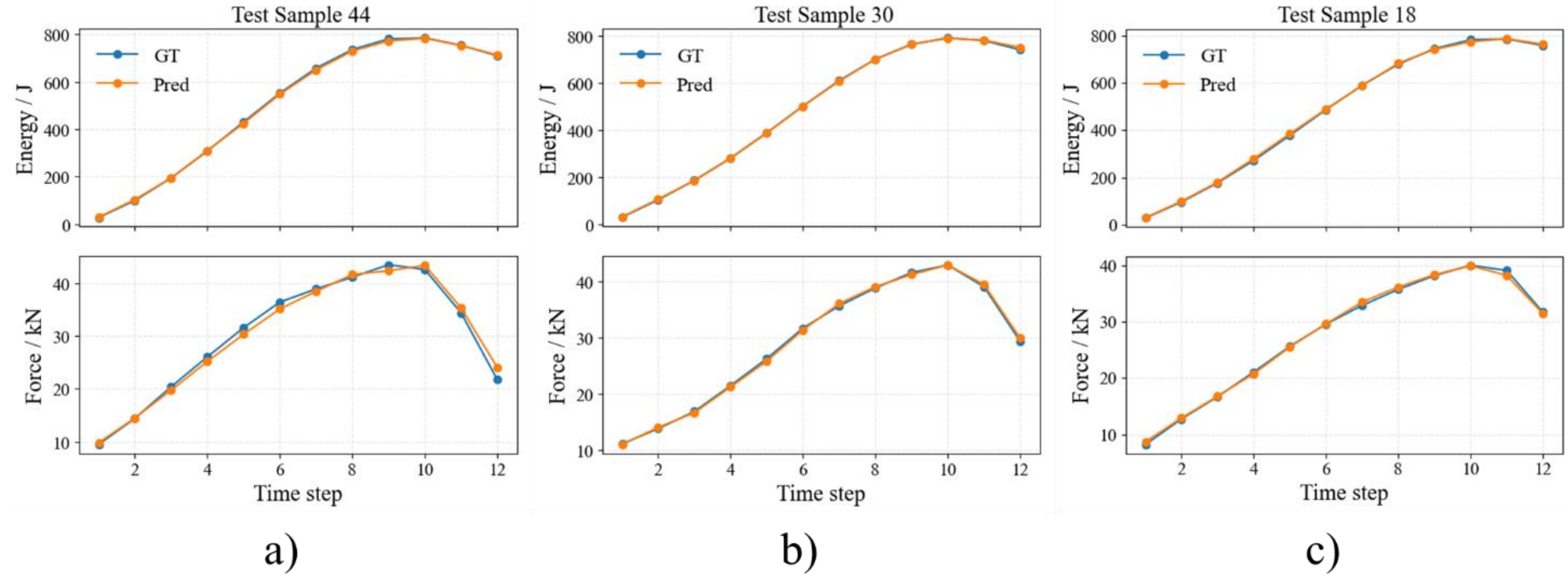

Figure 19: Comparison of predicted energy and force to the ground truth for all time steps of 3 representative test samples.

Finally, Table 7 reports summary error statistics for three engineering KPIs using relative error (in %) on the test set, including the bootstrap mean, standard deviation, and 95% bootstrap confidence interval. The tuned model achieves low average relative errors for maximum intrusion and especially internal energy, suggesting that energy absorption can be estimated reliably from the learned dynamics. The contact force exhibits higher uncertainty which is expected given that the overall magnitude of contact force is much smaller. The overall results demonstrate that ReGUNet maintains stable deformation rollouts while providing accurate and statistically well-supported predictions of both nodal displacements and key crashworthiness scalars over a broader design space.

Table 7: Error statistics for the engineering KPIs.

| Engineering KPIs | Mean test error (bootstrap) | Standard deviation of test error | 95% bootstrap CI of test error |
|---|---|---|---|
| Maximum intrusion / mm | 1.76% | 1.58% | 1.31% – 2.21% |
| Internal energy / J | 0.60% | 0.45% | 0.48% – 0.72% |
| Contact force / kN | 3.59% | 3.59% | 2.64% – 4.59% |

In addition to accuracy, inference time and memory consumption is evaluated to assess the practical efficiency of ReGUNet. On a single NVIDIA T4 GPU, inference of one complete sample (12 time steps) takes approximately 1.8 seconds, with a peak GPU memory usage of 2.5 GB. In contrast, the corresponding FE simulation requires around 5 minutes on a CPU workstation, yielding a 160× speed-up in total runtime. Additionally, the graph construction process, including mesh loading and graph generation, takes less than 1 second, making the overall surrogate pipeline substantially faster and lightweight compared to conventional FE workflows.

# 6 Conclusions

In this paper, a GNN-based surrogate model, Recurrent Graph U-Net (ReGUNet), is proposed to predict the crashworthiness performance of vehicle panel components. The model operates on graph-structured data, which is naturally a well-suited representation form of mesh data. To improve scalability and enable efficient information propagation between physically distant nodes, ReGUNet employs a hierarchical graph downsampling strategy by generating coarsened meshes of the underlying geometry. Multiple layers of coarsened graphs with different node densities are created to establish a U-Net architecture. Specialised downsampling and upsampling layers are employed for efficient and accurate message passing between graphs at different levels. At the finest and coarsest levels, recurrent message passing layers are utilised to enable message passing within the graphs in a recurrent manner. Message passing at the coarsest level can propagate feature information over a longer distance with fewer message passing steps, capturing spatial long-range dependencies between nodes. Each recurrent message passing layer not only performs information propagation within the graphs, but also propagates hidden states that contain historical information of previous time steps. This recurrence mechanism leads to more stabilised predictions over a larger number of time steps.

Using FE data of hot-stamped B-pillars under side impact, ReGUNet is shown to predict dynamic deformation behaviour with high accuracy. Comparisons against multiple baselines confirm that both hierarchical downsampling and recurrence contribute to improved long-horizon stability and reduced final-step deformation error, yielding more than 52% improvement in mean Euclidean distance relative to the baselines. In addition, an ablation study on cross-graph information propagation demonstrates that replacing DS-MP with DS-NW improves accuracy and robustness, leading to slower error growth over time and a narrower error distribution at the final time step. This highlights the significance of the proposed non-sharable weight downsampling strategy. Moreover, different training strategies are evaluated, where rollout training yields the best results for this study.

After hyperparameter tuning, ReGUNet demonstrates accurate and efficient prediction of the dynamic deformation behaviour of previously unseen B-pillar components across entire crash simulations. Furthermore, in a second case study with a wider design space, ReGUNet is extended to jointly predict node-level displacement fields and graph-level crashworthiness KPIs (internal energy and contact force), achieving strong agreement with FE results and low relative errors in key engineering metrics. Overall, these results indicate that ReGUNet provides a fast, reliable surrogate for component-level crashworthiness assessment.

Several promising potential directions remain for future work. First, the current experiments consider a relatively narrow design space, primarily driven by geometric variation, with fixed material properties, boundary and loading conditions. The framework can be extended by incorporating additional design variables, such as material parameters or wall thickness, as conditional inputs to the network. This could be achieved through dedicated feature-injection modules or transfer-learning strategies built on a pretrained backbone. Second, the hierarchical graph construction anchors coarse levels to a single benchmark mesh, which may bias cross-graph mappings when generalising to substantially different geometries. More adaptive and generalisable downsampling strategies will be further investigated. In addition, while the present approach is purely data-driven, future work will explore physics-aware enhancements to improve physical consistency, for example by introducing penalty terms for non-physical node interpenetration or regularisation based on global energy consistency. Future research directions also include exploring the model's potential in multi-component simulations, by incorporating varying material properties and more complex boundary conditions. Eventually, the approach could be applied to full-vehicle crash simulations, providing more comprehensive and realistic assessments.

## Author contributions (CRediT):

Haoran Li: Conceptualisation, Data curation, Formal analysis, Investigation, Methodology, Validation, Visualization, Writing – original draft. Yingxue Zhao: Conceptualisation, Methodology, Validation, Writing – review & editing. Haosu Zhou: Conceptualisation, Validation, Writing – review & editing. Tobias Pfaff: Validation, Writing – review & editing. Nan Li: Conceptualization, Project administration, Supervision, Validation, Writing – review & editing.

## Data availability

The data that support the findings of this study will be made available from the corresponding author upon reasonable request.

## Acknowledgements

The authors acknowledge funding support from UKRI (UKRI221: AI-Driven Design for Forming High-Performance Vehicle Parts), as well as PhD scholarships from Imperial College London. They would also like to thank ESI Group for providing technical support with the Virtual-Performance Solution (VPS). The authors would like to specifically acknowledge Mr. Ziane Mustapha from ESI Group for his valuable assistance with the simulation setup. For the purpose of open access, the authors have applied a Creative Commons Attribution (CC BY) license to any Author Accepted Manuscript version arising.

## References

1. Kim, H.C. and T.J. Wallington, *Life-Cycle Energy and Greenhouse Gas Emission Benefits of Lightweighting in Automobiles: Review and Harmonization.* Environmental Science & Technology, 2013. **47**(12): p. 6089-6097.
2. IIHS, *Side Impact Crashworthiness Evaluation 2.0 Crash Test Protocol*. 2022, Insurance Institute for Highway Safety.
3. Rogala, M., J. Gajewski, and M. Ferdynus, *The Effect of Geometrical Non-Linearity on the Crashworthiness of Thin-Walled Conical Energy-Absorbers.* Materials, 2020. **13**(21): p. 4857.
4. Sakaridis, E., N. Karathanasopoulos, and D. Mohr, *Machine-learning based prediction of crash response of tubular structures.* International Journal of Impact Engineering, 2022. **166**.
5. Albak, E.İ., *Optimization design for circular multi-cell thin-walled tubes with discrete and continuous design variables.* Mechanics of Advanced Materials and Structures, 2023. **30**(24): p. 5091-5105.
6. Ahmadi Dastjerdi, A., et al., *Crushing analysis and multi-objective optimization of different length bi-thin walled cylindrical structures under axial impact loading.* Engineering Optimization, 2019. **51**: p. 1-18.
7. Zende, P. and H. Dalir. *Multi-Objective Optimization of Composite Square Tube for Minimizing Peak Crushing Force and Maximizing Specific Energy Absorption Using Artificial Neural Network and Genetic Algorithm*. in *ASME International Mechanical Engineering Congress and Exposition, Proceedings (IMECE)*. 2022.
8. Xiong, F., et al., *Multi-objective lightweight and crashworthiness optimization for the side structure of an automobile body.* Structural and Multidisciplinary Optimization, 2018. **58**(4): p. 1823-1843.
9. Kohar, C.P., et al., *Using Artificial Intelligence to Aid Vehicle Lightweighting in Crashworthiness with Aluminum.* MATEC Web Conf., 2020. **326**: p. 01006.
10. Guo, W., et al., *Machine learning-based crashworthiness optimization for the square cone energy-absorbing structure of the subway vehicle.* Structural and Multidisciplinary Optimization, 2023. **66**(8): p. 182.
11. Kohar, C.P., et al., *A machine learning framework for accelerating the design process using CAE simulations: An application to finite element analysis in structural crashworthiness.* Computer Methods in Applied Mechanics and Engineering, 2021. **385**.
12. Li, H., H. Zhou, and N. Li, *An integrated convolutional neural network-based surrogate model for crashworthiness performance prediction of hot-stamped vehicle panel components.* MATEC Web of Conferences, 2024. **401**.
13. Wen, Z., et al., *Data-driven spatiotemporal modeling for structural dynamics on irregular domains by stochastic dependency neural estimation.* Computer Methods in Applied Mechanics and Engineering, 2023. **404**: p. 115831.
14. Deshpande, S., J. Lengiewicz, and S. Bordas, *MAgNET: A Graph U-Net Architecture for Mesh-Based Simulations.* arXiv preprint arXiv:2211.00713, 2022.
15. He, J., et al., *On the use of graph neural networks and shape-function-based gradient computation in the deep energy method.* International Journal for Numerical Methods in Engineering, 2023. **124**(4): p. 864-879.
16. Fu, X., et al., *An finite element analysis surrogate model with boundary oriented graph embedding approach for rapid design.* Journal of Computational Design and Engineering, 2023. **10**(3): p. 1026-1046.

17. Chen, Q., et al., *Predicting dynamic responses of continuous deformable bodies:A graph-based learning approach.* Computer Methods in Applied Mechanics and Engineering, 2024. **420**: p. 116669.

18. Jin, Z., et al., *Leveraging graph neural networks and neural operator techniques for high-fidelity mesh-based physics simulations.* APL Machine Learning, 2023. **1**(4): p. 046109.

19. Chou, Y.-T., et al., *StructGNN: An efficient graph neural network framework for static structural analysis.* Computers & Structures, 2024. **299**: p. 107385.

20. Parisi, F., et al., *On the use of mechanics-informed models to structural engineering systems: Application of graph neural networks for structural analysis.* Structures, 2024. **59**: p. 105712.

21. Liu, Q., et al., *Fluid Simulation System Based on Graph Neural Network.* arXiv preprint arXiv:2202.12619, 2022.

22. He, X., Y. Wang, and J. Li, *Flow completion network: Inferring the fluid dynamics from incomplete flow information using graph neural networks.* Physics of Fluids, 2022. **34**(8): p. 087114.

23. Yang, Z., et al., *AMGNET: multi-scale graph neural networks for flow field prediction.* Connection Science, 2022. **34**(1): p. 2500-2519.

24. Li, T., et al., *Finite Volume Graph Network (FVGN): Predicting unsteady incompressible fluid dynamics with finite volume informed neural network.* arXiv preprint arXiv:2309.10050, 2023.

25. Xue, T., S. Adriaenssens, and S. Mao, *Learning the nonlinear dynamics of mechanical metamaterials with graph networks.* International Journal of Mechanical Sciences, 2023. **238**: p. 107835.

26. Meyer, P.P., et al., *Graph-based metamaterials: Deep learning of structure-property relations.* Materials & Design, 2022. **223**: p. 111175.

27. Pfaff, T., et al., *Learning mesh-based simulation with graph networks.* arXiv preprint arXiv:2010.03409, 2020.

28. Cao, Y., et al., *Efficient Learning of Mesh-Based Physical Simulation with BSMS-GNN.* arXiv preprint arXiv:2210.02573, 2022.

29. Fortunato, M., et al., *Multiscale meshgraphnets*, in *2nd AI4Science Workshop at the 39th International Conference on Machine Learning (ICML), 2022*. 2022.

30. Lea, C., et al. *Temporal convolutional networks: A unified approach to action segmentation*. in *Computer Vision–ECCV 2016 Workshops: Amsterdam, The Netherlands, October 8-10 and 15-16, 2016, Proceedings, Part III 14*. 2016. Springer.

31. Kipf, T.N. and M. Welling, *Semi-supervised classification with graph convolutional networks.* arXiv preprint arXiv:1609.02907, 2016.

32. Zhao, Y., et al., *A review of graph neural network applications in mechanics-related domains.* Artificial Intelligence Review, 2024. **57**.

33. Sanchez-Gonzalez, A., et al., *Learning to Simulate Complex Physics with Graph Networks*, in *Proceedings of the 37th International Conference on Machine Learning*, D. Hal, III and S. Aarti, Editors. 2020, PMLR: Proceedings of Machine Learning Research. p. 8459--8468.

34. Gilmer, J., et al., *Neural Message Passing for Quantum Chemistry*, in *Proceedings of the 34th International Conference on Machine Learning*, P. Doina and T. Yee Whye, Editors. 2017, PMLR: Proceedings of Machine Learning Research. p. 1263--1272.

35. Lino, M., et al., *Simulating continuum mechanics with multi-scale graph neural networks.* arXiv preprint arXiv:2106.04900, 2021.

36. Liu, W., M. Yagoubi, and M. Schoenauer. *Multi-resolution Graph Neural Networks for PDE Approximation*. in *Artificial Neural Networks and Machine Learning – ICANN 2021*. 2021. Cham: Springer International Publishing.

37. Lino, M., et al., *Towards fast simulation of environmental fluid mechanics with multi-scale graph neural networks.* arXiv preprint arXiv:2205.02637, 2022.

38. Lino, M., et al., *Multi-scale rotation-equivariant graph neural networks for unsteady Eulerian fluid dynamics.* Physics of Fluids, 2022. **34**(8): p. 087110.

39. Brumand-Poor, F., M. Trautmann, and K. Schmitz, *Advancing deformation calculation: a physics-informed deep graph learning framework for hyperelastic materials.* Advanced Modeling and Simulation in Engineering Sciences, 2025. **12**(1): p. 20.

40. Dalton, D., D. Husmeier, and H. Gao, *Physics-informed graph neural network emulation of soft-tissue mechanics.* Computer Methods in Applied Mechanics and Engineering, 2023. **417**: p. 116351.

41. Valente, M., et al., *Physics-consistent machine learning with output projection onto physical manifolds.* Communications Physics, 2025. **8**(1): p. 433.

42. Li, X., et al., *Development of surrogate models in reliability-based design optimization: A review.* Mathematical Biosciences and Engineering, 2021. **18**: p. 6386-6409.

43. Qiu, N., et al., *Crashworthiness optimization with uncertainty from surrogate model and numerical error.* Thin-Walled Structures, 2018. **129**: p. 457-472.

44. Zhang, X., J. Zhou, and Z. Feng, *B-pillar collision test method*. 2020, Aiways Automobile Shanghai Co Ltd: China.

45. Li, N., *Fundamentals of Materials Modelling for Hot Stamping of UHSS Panels with Graded Properties.* Imperial College London, 2013.

46. He, K., et al. *Deep Residual Learning for Image Recognition*. in *2016 IEEE Conference on Computer Vision and Pattern Recognition (CVPR)*. 2016.

47. Bengio, S., et al., *Scheduled sampling for sequence prediction with recurrent Neural networks*, in *Proceedings of the 29th International Conference on Neural Information Processing Systems - Volume 1*. 2015, MIT Press: Montreal, Canada. p. 1171–1179.

# Appendix A

To demonstrate the diversity of the dataset, the nearest-neighbour distance to the training set in a normalised design-parameter space is computed for each validation sample $i$. Each raw design entry is stored as a triplet $(x_i, c_i, z_i)$, where $x_i$ is the global x-morph parameter, $c_i \in \{0,1,2\}$ indicates which z-control point is active, and $z_i$ is the corresponding z-morph magnitude. This triplet is converted into a 4D design vector,

$$\mathbf{d}_i = \begin{bmatrix} x_i \\ \Delta z_{i,1} \\ \Delta z_{i,2} \\ \Delta z_{i,3} \end{bmatrix}, \text{with } \Delta z_{i,k} = \begin{cases} z_i, & k = c_i + 1, \\ 0, & otherwise. \end{cases} \tag{7}$$

i.e., only one of the three $z$-entries is non-zero for each sample.

To make distances comparable across dimensions, each component of $\mathbf{d}_i$ is normalised (via elementwise division) by the maximum absolute value across the combined training and validation sets to get $\hat{\mathbf{d}}_i$. For each validation sample $i \in \mathcal{V}$, its Euclidean distance to every training sample $j \in \mathcal{T}$ in the normalised space, and take the minimum:

$$\delta_j = \min_{i \in \mathcal{V}} \left\| \hat{\mathbf{d}}_i - \hat{\mathbf{d}}_j \right\|_2 \tag{8}$$

This distance indicates the similarity of the validation sample to the most similar training sample, where a larger mean distance indicates a more diverse dataset. Using this definition, the mean nearest-neighbour distances from validation to training is 0.1348. This indicates that on average, each validation design is separated from its closest training design by approximately 13.5% in this normalised space. This suggests that the validation set is not composed of trivial near-duplicates of training samples under the adopted design-parameter representation and distance metric.

## Appendix B

To assess the robustness of ReGUNet to stochastic effects in optimisation, we repeat training and evaluation using 4 independent random seeds (15, 21, 36, 42). All runs use the same datasets in Case study 1, tuned ReGUNet architecture, and training protocol, and are evaluated under autoregressive rollout after teacher forcing training. Performance is quantified by the average mean Euclidean distance (in mm) between predicted and ground-truth node at each time step. In addition to the time-resolved error curves, we summarise the final-timestep error distribution and report bootstrap 95% confidence intervals (CIs) for the mean final-timestep error.

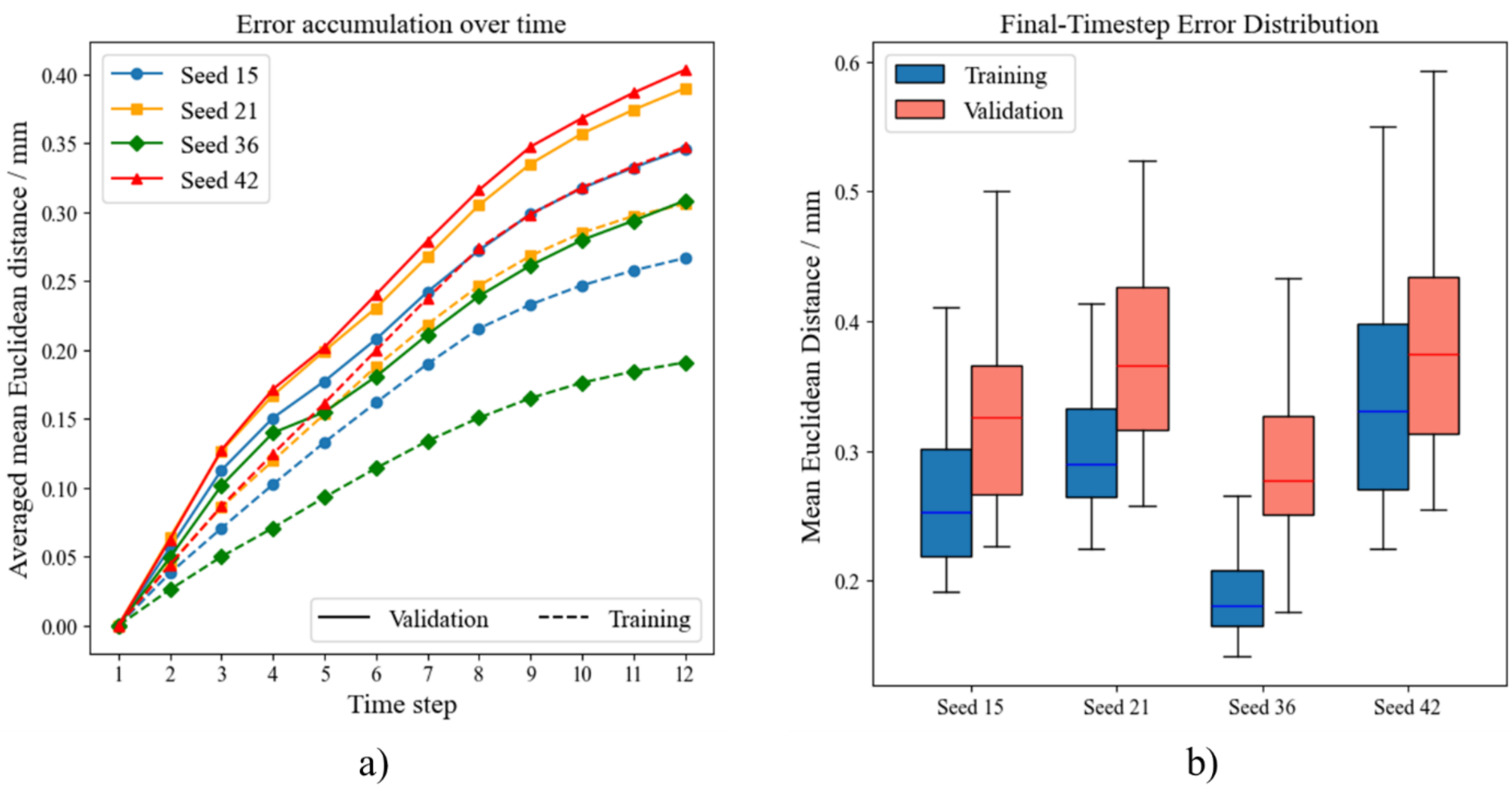

Figure 20: Comparisons between the 4 random seeds in terms of
a) error accumulation over 12 time steps, and b) final time-step error distributions, in terms of mean Euclidean distance.

Figure 20 a) shows the error accumulation over time for both training and validation sets. Across all seeds, the rollout error increases monotonically with time step, consistent with the expected compounding of small prediction errors under autoregressive inference. Importantly, the hyperparameter tuning is performed using seed 36. As a result, it is expected that seed 36 exhibits comparatively stronger performance, and this is reflected in both Figure 20 a) and b), where seed 36 consistently achieves the lowest error trajectory and the smallest final-timestep mean error. While the curves diverge at later time steps, the overall spread remains moderate in absolute terms. From Table 8, one can see that the final-timestep mean errors range from 0.31 mm (seed 36) to 0.40 mm (seed 42), with comparable variability (std $\approx$ 0.12 – 0.15 mm) and relatively tight bootstrap 95% confidence intervals for the mean. Although the separation between curves can appear visually pronounced when plotted across time, this variation is small relative to the performance gaps observed between different baseline models in earlier comparisons.

Table 8: The effect of random seed difference on model performance.

| Random seed | Mean validation error (bootstrap) / mm | Standard deviation of validation error / mm | 95% bootstrap CI of validation error / mm |
|---|---|---|---|
| 15 | 0.35 | 0.15 | 0.30 – 0.39 |
| 21 | 0.39 | 0.13 | 0.35 – 0.43 |
| 36 | 0.31 | 0.12 | 0.27 – 0.34 |
| 42 | 0.40 | 0.14 | 0.36 – 0.44 |

More importantly, all seeds achieve improved accuracy compared to the untuned model, both in the time-resolved accumulation curves and in the final-timestep error distributions. In other words, even accounting for stochasticity, the tuned configuration yields consistently better rollout performance across independent runs, while maintaining stable training/validation behaviour. This supports the conclusion that the reported gains are not driven by a single favourable seed, but are reproducible across multiple training runs.

# Appendix C

The full dynamic prediction of a representative test sample in Case study 1 is shown in this appendix. For each time step, the ground-truth z-displacement, the predicted z-displacement, and the corresponding error map are shown as nodal contour plots. Overall, the predicted deformation fields exhibit good agreement with the ground truth across all time steps.

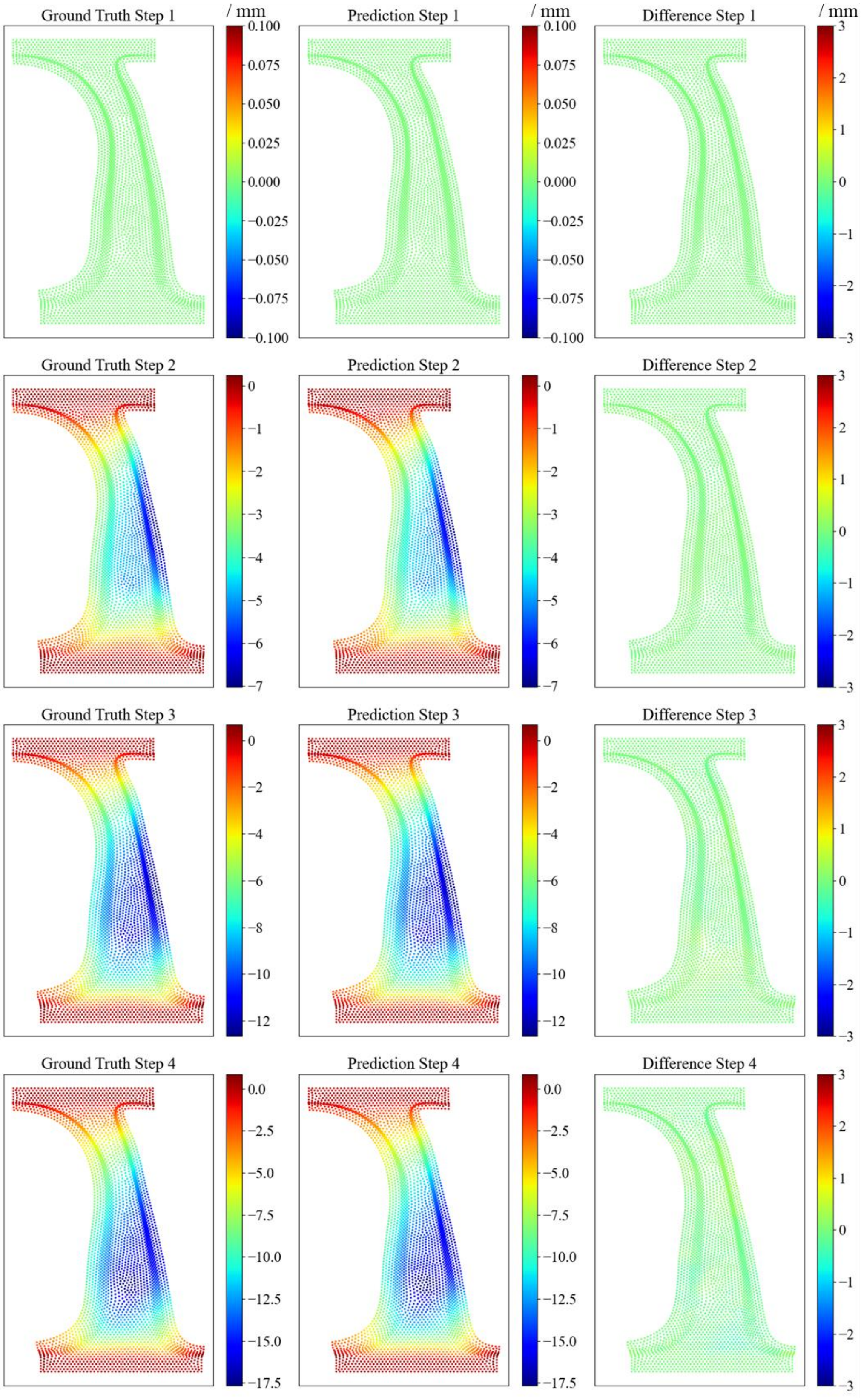

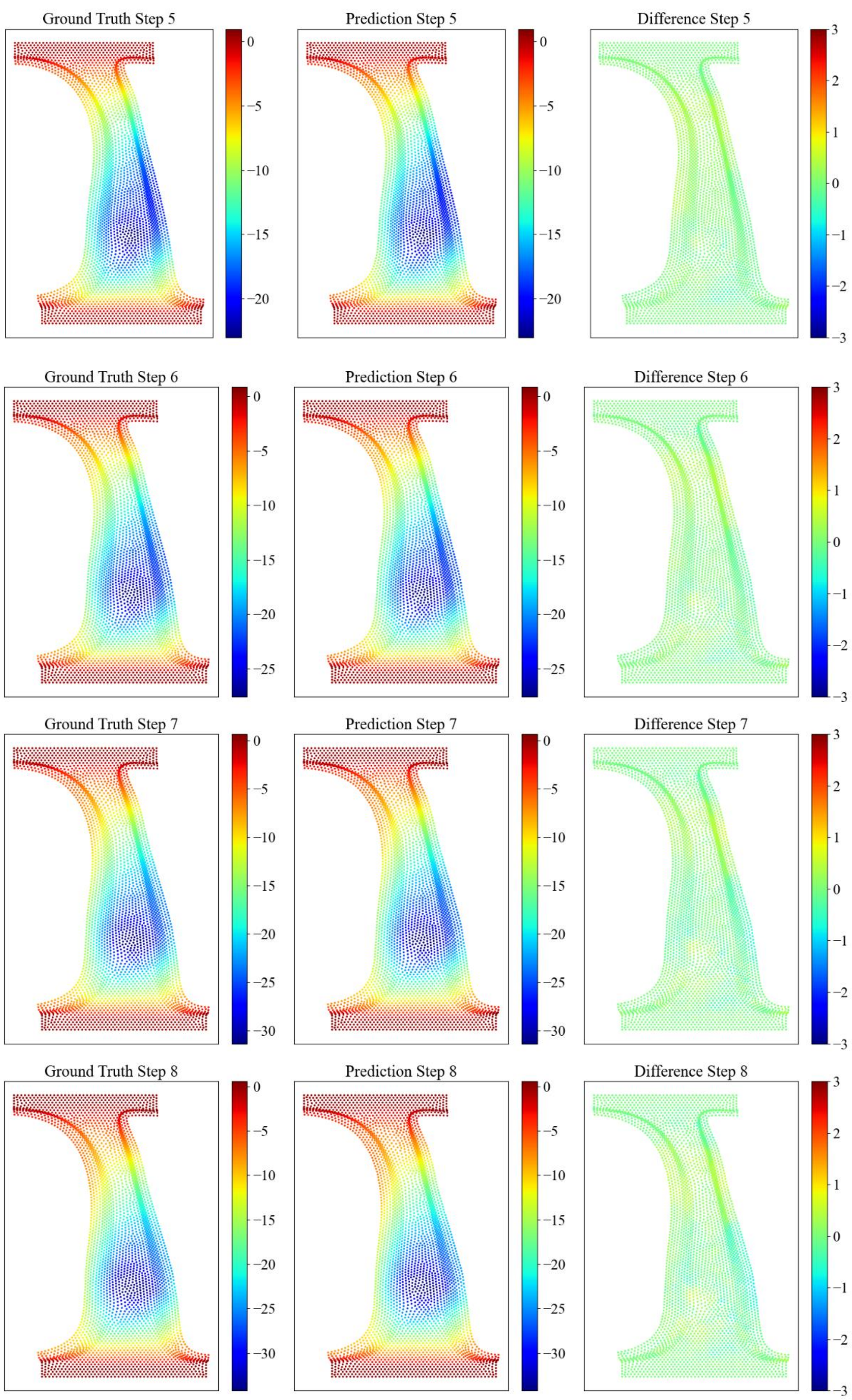

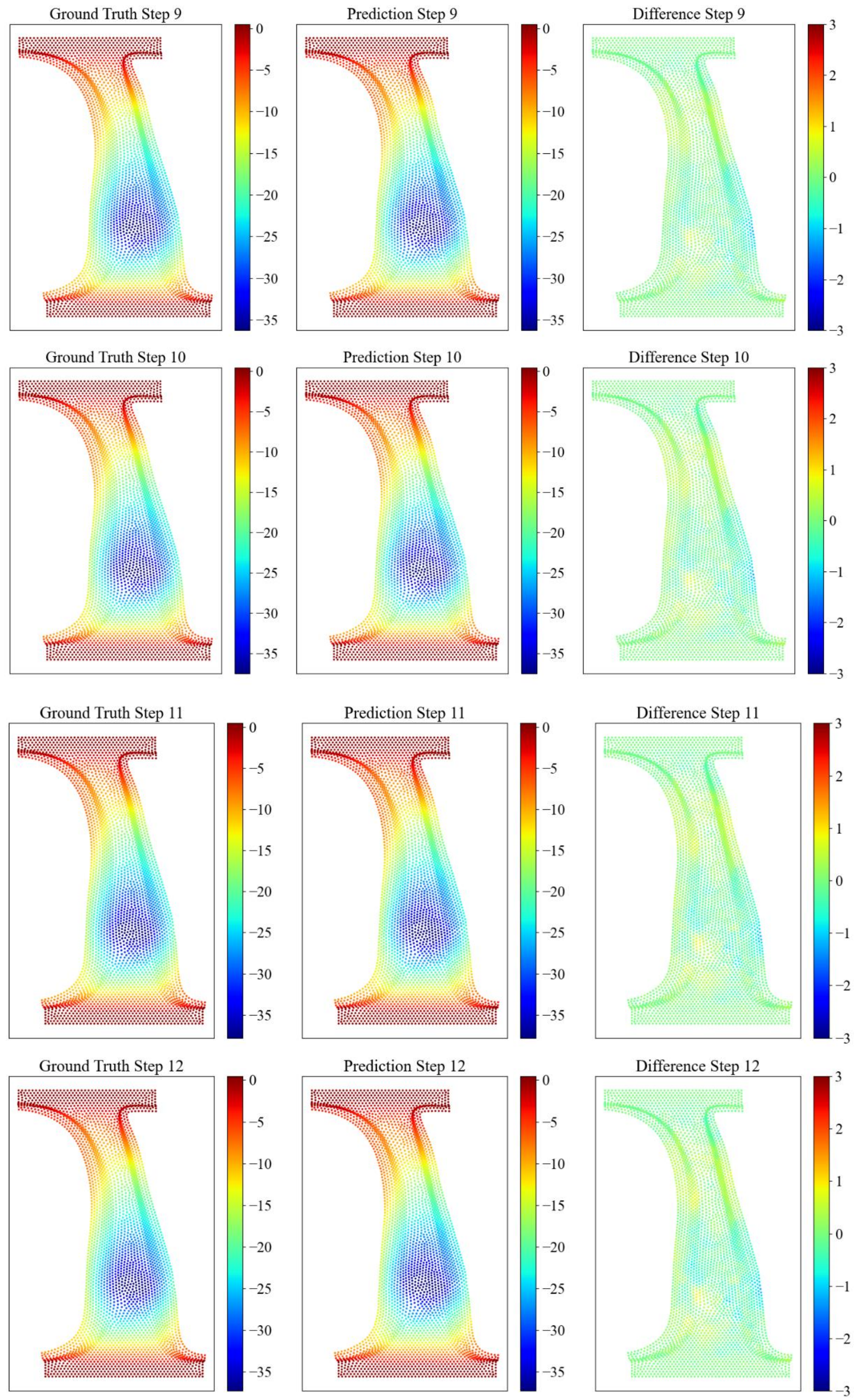

Figure 21: Full dynamic prediction of z-displacement of a representative test sample B-pillar.

# Appendix D

Appendix D provides a finer-grained view of the scalar-output performance by reporting predicted vs. ground truth scatter plots at every time step (t = 1…12) on the test set, for (i) internal energy (J) and (ii) contact force (kN). Each subplot includes the corresponding $R^2$ value, allowing the temporal evolution of prediction fidelity to be inspected directly rather than only at the final step.

Overall, the points cluster closely around the $y = x$ line for most time steps, indicating good agreement. The contact force prediction remains consistently strong across the trajectory, with most values staying in a high range. The internal energy prediction is also accurate for the majority of steps. Worth noting, the lower $R^2$ observed at early time steps and at $t = 10$ does not necessarily indicate poor prediction quality, but rather reflects the fact that the ground truth values are concentrated within a narrow range, so most samples lie close to their mean and $R^2$ becomes less informative.

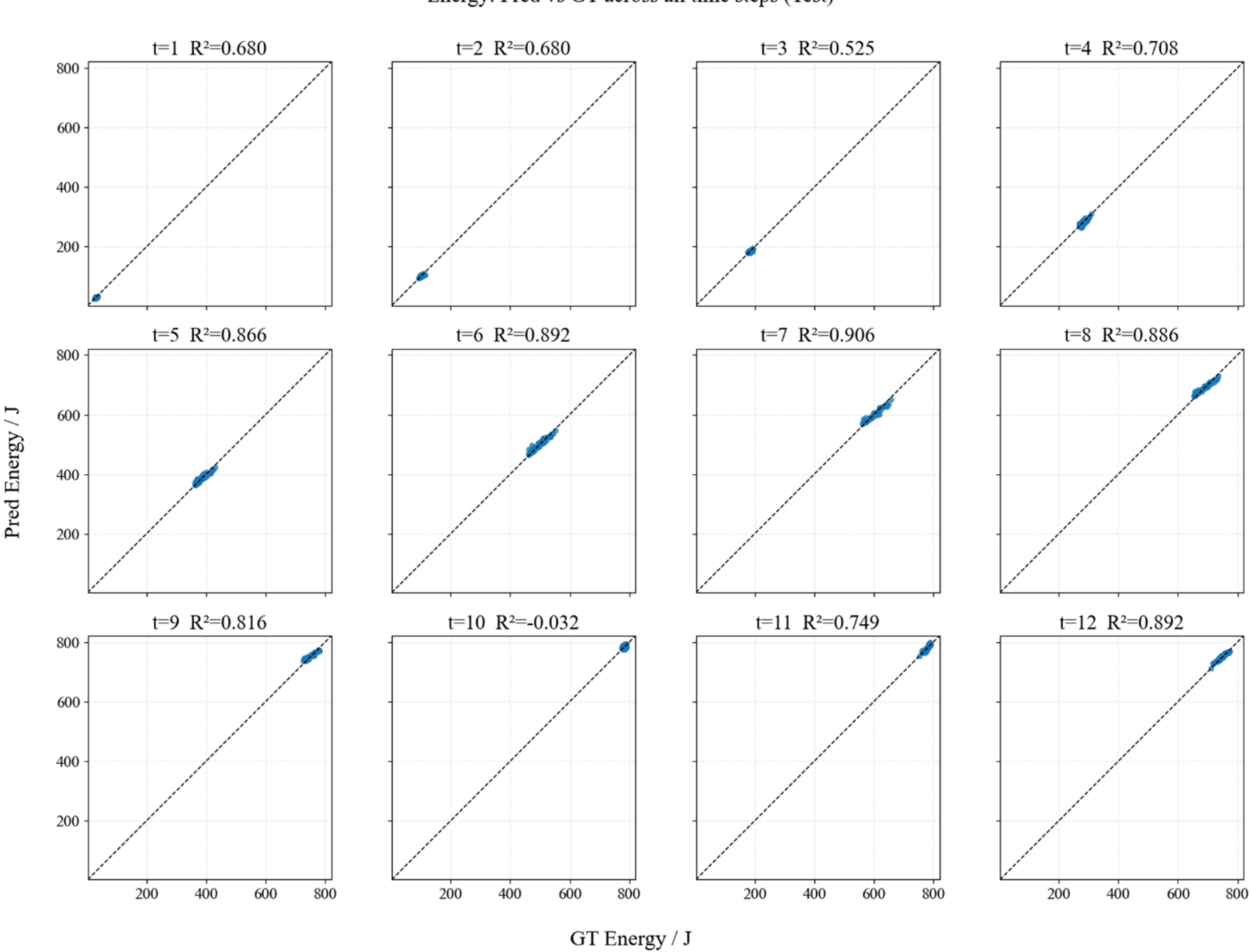

Figure 22: Predicted vs. ground-truth scatter plots of internal energy for the test set across all 12 time steps

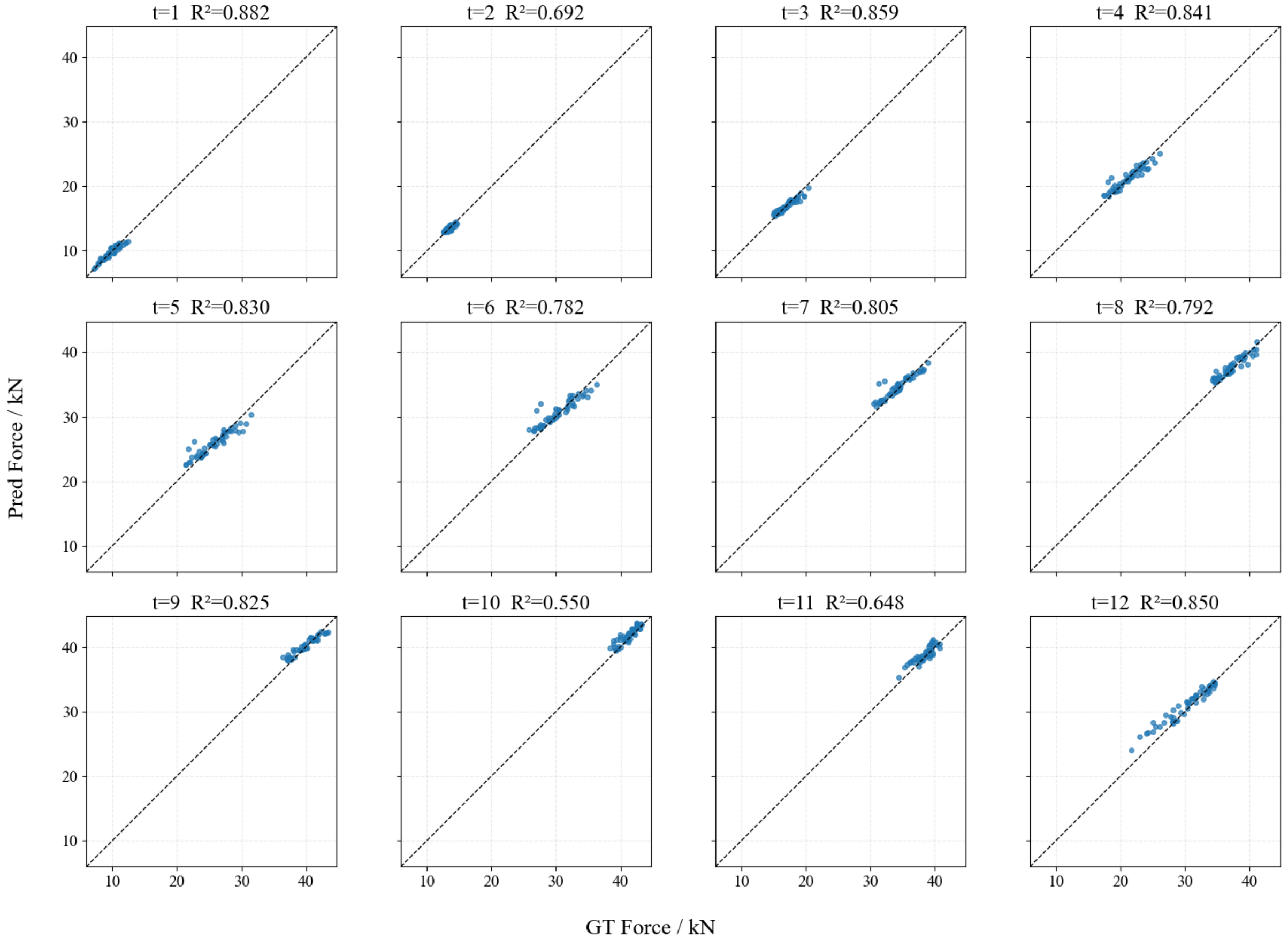

Figure 23: Predicted vs. ground-truth scatter plot for contact force for the test set across all 12 time steps.